\documentclass[%
pre,reprint,
superscriptaddress,
amsmath,amssymb,
aps,
]{revtex4-1}

\usepackage{graphicx}% Include figure files
\usepackage{dcolumn}% Align table columns on decimal point
\usepackage{bm}% bold math
\usepackage{textcomp}
\usepackage[pdftex,colorlinks=true,bookmarks=false,citecolor=blue,urlcolor=blue]{hyperref}
\usepackage{cancel}
\usepackage{color}
\newcolumntype{?}{!{\vrule width 1.1pt}}
\allowdisplaybreaks
\newcommand{\mrm}[1]{\mathrm{#1}}

\newcommand{\rmi}{\mathrm{i}} % imaginary unit
\newcommand{\Qvec}{\mathbf{Q}}
\newcommand{\Svec}{\mathbf{S}}	%Poynting vektor
\newcommand{\Evec}{\mathbf{E}}	%el.Feld
\newcommand{\Bvec}{\mathbf{B}}	%mag.Induktion
\newcommand{\Jvec}{\mathbf{J}}	%Strom
\newcommand{\rvec}{\mathbf{r}}
\newcommand{\evec}{\mathbf{e}}	%el.Feld
\newcommand{\first}{$1^{\text{st}}\,$}
\newcommand{\second}{$2^{\text{nd}}\,$}

\DeclareSymbolFont{lettersA}{U}{txmia}{m}{it}
\DeclareMathSymbol{\real}{\mathord}{lettersA}{"92} % real numbers
\DeclareMathSymbol{\cplx}{\mathord}{lettersA}{"83} % complex numbers

\newcommand{\change}[1]{{\textcolor{black}{#1}}}
\newcommand{\rev}[1]{{\color{black}#1}}
\newcommand{\blue}[1]{{\color{black}#1}}

\begin{document}

%\preprint{APS/123-QED}

\title{Terahertz emission from laser-driven gas-plasmas: a plasmonic point of view}
%\thanks{A footnote to the article title}%

\author{I.~Thiele}
\email{illia-thiele@web.de}
\affiliation{Univ.~Bordeaux - CNRS - CEA, Centre Lasers Intenses et Applications, UMR 5107, 33405 Talence, France}
\affiliation{Department of Physics, Chalmers University of Technology, SE-412 96 G{\"o}teborg, Sweden}

\author{B.~Zhou}
\affiliation{DTU Fotonik—Department of Photonics Engineering, Technical University of Denmark, DK-2800 Kongens Lyngby, Denmark}

\author{A.~Nguyen}
\affiliation{CEA/DAM {\^I}le-de-France, Bruy\`eres-le-Ch\^atel, 91297 Arpajon, France}

\author{E.~Smetanina}
\affiliation{Univ.~Bordeaux - CNRS - CEA, Centre Lasers Intenses et Applications, UMR 5107, 33405 Talence, France}
\affiliation{Department of Physics, University of Gothenburg, SE-412 96 G{\"o}teborg, Sweden}

\author{R.~Nuter}
\affiliation{Univ.~Bordeaux - CNRS - CEA, Centre Lasers Intenses et Applications, UMR 5107, 33405 Talence, France}

\author{K.~J.~Kaltenecker}
\affiliation{DTU Fotonik—Department of Photonics Engineering, Technical University of Denmark, DK-2800 Kongens Lyngby, Denmark}

\author{J.~D\'echard}
\affiliation{CEA/DAM {\^I}le-de-France, Bruy\`eres-le-Ch\^atel, 91297 Arpajon, France}

\author{P.~Gonz{\'a}lez de Alaiza Mart{\'i}nez}
\affiliation{Univ.~Bordeaux - CNRS - CEA, Centre Lasers Intenses et Applications, UMR 5107, 33405 Talence, France}

\author{L.~Berg\'e}
\affiliation{CEA/DAM {\^I}le-de-France, Bruy\`eres-le-Ch\^atel, 91297 Arpajon, France}

\author{P.~U.~Jepsen}
\affiliation{DTU Fotonik—Department of Photonics Engineering, Technical University of Denmark, DK-2800 Kongens Lyngby, Denmark}

\author{S.~Skupin}
\affiliation{Univ.~Bordeaux - CNRS - CEA, Centre Lasers Intenses et Applications, UMR 5107, 33405 Talence, France}
\affiliation{Institut Lumi\`ere Mati\`ere, UMR 5306 Universit\'e Lyon 1 - CNRS, Universit\'e de Lyon, 69622 Villeurbanne, France}

\date{\today}

\begin{abstract}
	We disclose an unanticipated link between plasmonics and nonlinear frequency down-conversion in laser-induced gas-plasmas.
	For two-color femtosecond pump pulses, a plasmonic resonance is
	shown to broaden the terahertz emission spectra significantly. We identify the resonance as a leaky mode, which contributes to the emission spectra whenever electrons are excited along a direction where the plasma size is smaller than the plasma wavelength.
	As a direct consequence, such resonances can be \change{controlled} by changing the polarization properties of elliptically-shaped driving \change{laser pulses}.
	Both\change{, experimental results and} 3D Maxwell consistent simulations confirm that a significant terahertz pulse shortening and spectral broadening can be achieved by exploiting the transverse driving \change{laser beam} shape as an additional degree of freedom. 
\end{abstract}
	
	\maketitle
	
	%%%%%%%%%%%%%%%%%%%%%%%%%%  body  %%%%%%%%%%%%%%%%%%%%%%%%%%
	\section{Introduction}
	
	Terahertz~(THz) radiation has become an ubiquitous tool for many applications in science and technology~\cite{Kampfrath,Tonouchi}. Quite a number of those applications, as for example THz time-domain spectroscopy, require broadband THz sources. Unlike conventional THz sources such as photo-conductive switches~\cite{Tonouchi} or quantum cascade lasers~\cite{Vitiello:15}, laser-induced gas-plasmas straightforwardly produce emission from THz up to far-infrared frequencies~\cite{Kim}. In the standard setup, a femtosecond (fs) two-color~(2C) laser pulse composed of fundamental harmonic~(FH) and second harmonic~\change{(SH)} frequency is focused into an initially neutral gas creating free electrons via tunnel ionization. 
	These electrons are accelerated by the laser electric field and produce a macroscopic current leading to broadband THz emission.
	
	Numerous experimental results show that the laser-induced free electron density has a strong impact on the THz emission spectra~\cite{Hamster94,Kim,PhysRevLett.105.053903,Li:16,PhysRevLett.116.063902}. While it is frequently observed that a larger free electron density leads to broader THz spectra, the origin of the effect remains controversial. In \cite{Hamster94,Li:16,PhysRevLett.116.063902}, homogeneous plasma oscillations were proposed as an explanation, even though those oscillations are in principle non-radiative~\cite{PhysRevLett.89.209301,Bystrov2005,Gildenburg07,PhysRevE.94.063202,PhysRevE.69.066415,Gorbunov06,PedrosArticle,doi:10.1063/1.4953098}. 
	Moreover, nonlinear propagation effects have been held responsible for THz spectral broadening as well~\cite{PhysRevLett.105.053903,CabreraGranado2015}.
	%Nevertheless, the conversion of ponderomotively excited plasma oscillations to radiating modes has already been predicted~\cite{PhysRevE.69.066415,Gorbunov06} and confirmed by particle-in-cell simulations~\cite{PedrosArticle,doi:10.1063/1.4953098}, where besides the electron density also the strength of the electron density gradient has been pointed out as a crucial factor. 
	
	\begin{figure}[!t]
		\centering
		\includegraphics[width=0.49\columnwidth]{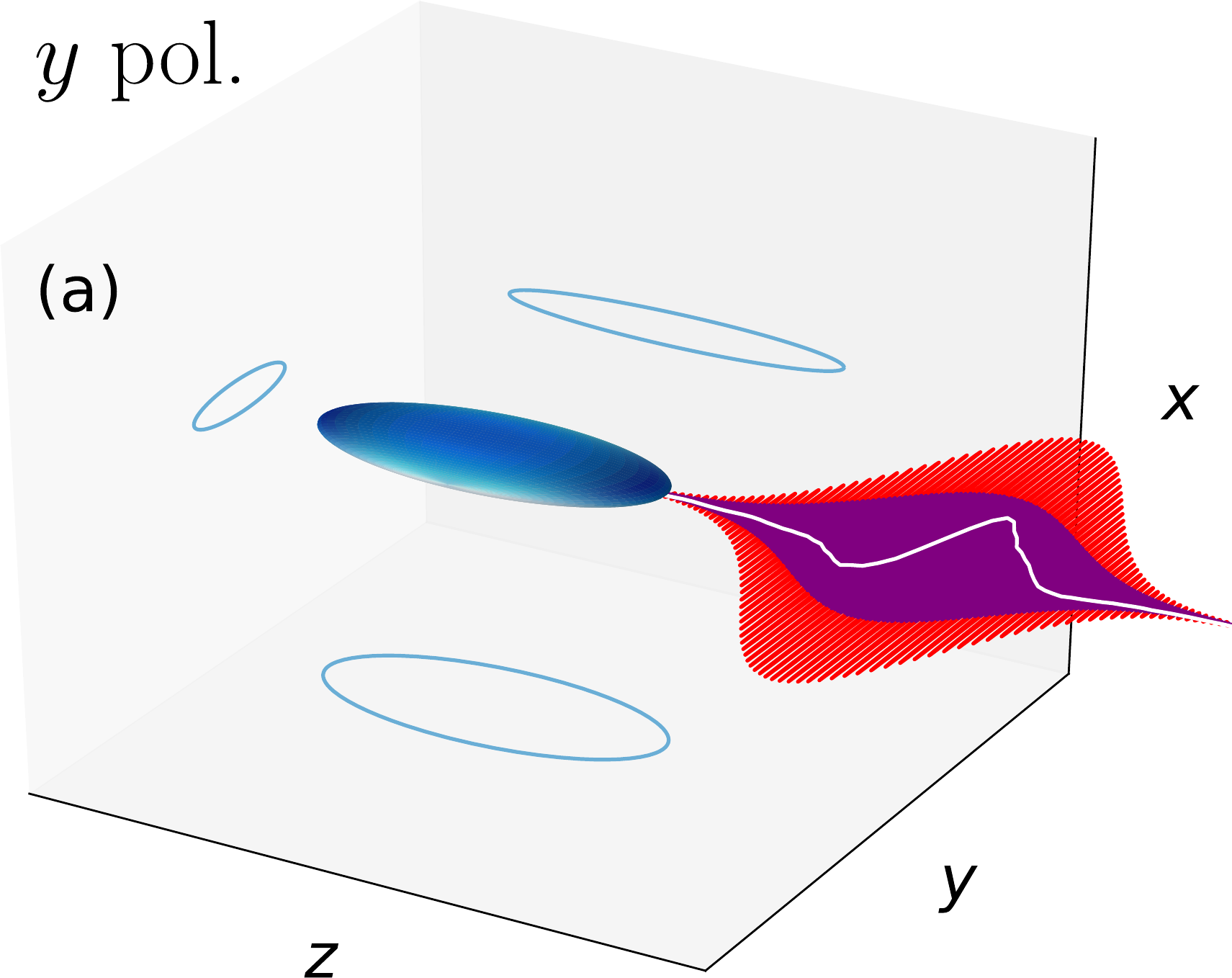}
		\includegraphics[width=0.49\columnwidth]{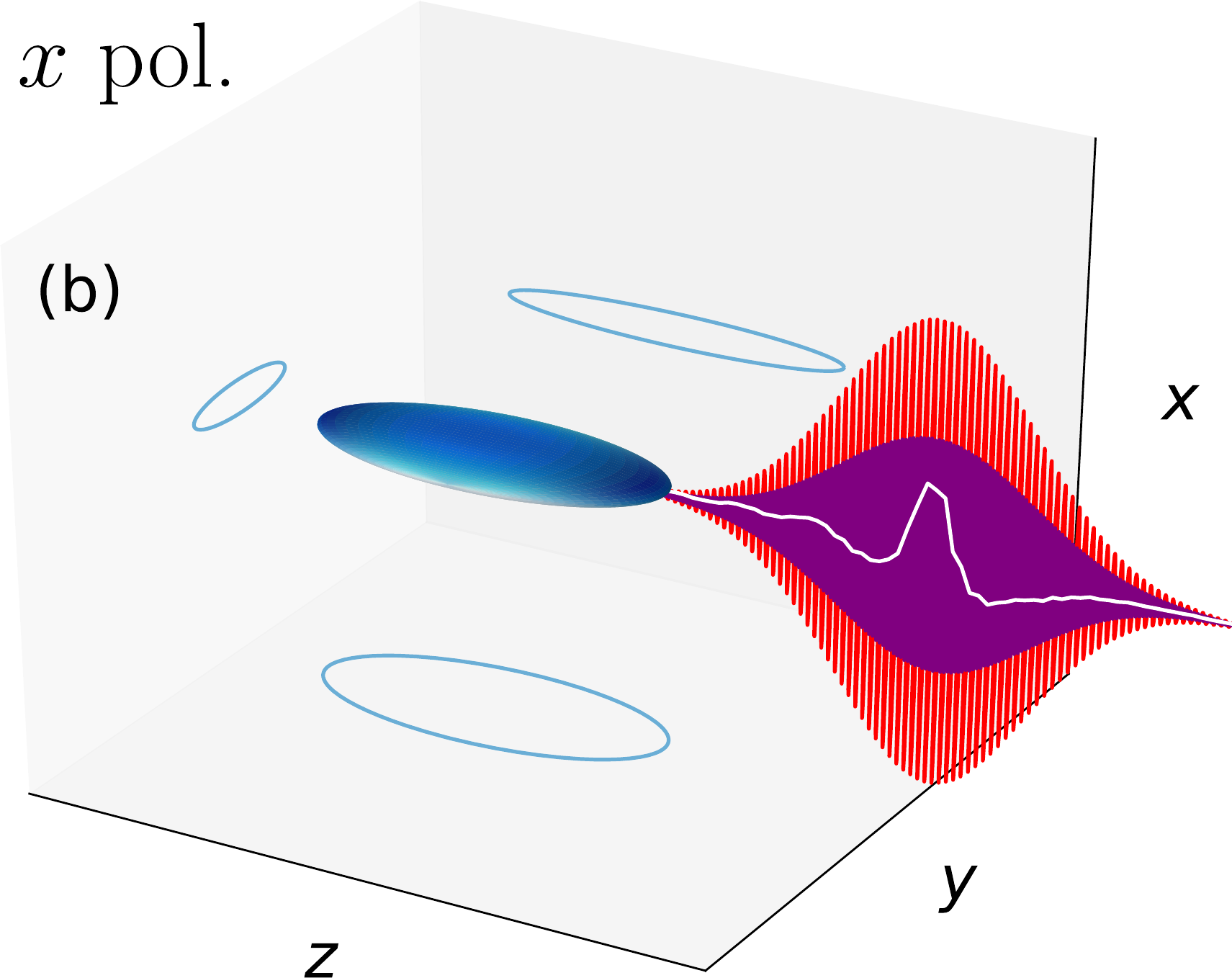}
		\caption{Illustrated configurations of THz emission from an ellipsoidal plasma induced by a 2C Gaussian laser pulse~(FH in red, SH in purple) with strongly elliptical beam shape propagating along $z$. The laser electric field is $y$-polarized \blue{(along the long axis of the elliptical beam)} in (a) and $x$-polarized \blue{(along the short axis of the elliptical beam)} in (b). The plasma is sketched \change{as a blue tri-axial ellipsoid, and its projections are shown in the respective planes.} \change{Experimentally measured forward} emitted THz pulses are presented as white lines demonstrating a significantly shorter pulse duration for \blue{an $x$-polarized pulse}, which can be attributed to triggering a plasmonic resonance~(see Sec.~\ref{sec:theo} for details).}
		\label{fig:scheme}
	\end{figure}
	
	On the other hand, the gas plasma produced by the fs laser pulse is a finite conducting structure with a lifetime largely exceeding the fs time scale. 
	Thus, one can expect that the gas plasma features plasmonic resonances which may have a \change{significant} impact on the THz emission properties~\cite{Kostin:10,Kostin15}.
	However, no direct evidence of plasmonic effects in laser-induced gas-plasmas was observed so far\change{, because it is hard to distinguish them in experiments.}
	Also from the theoretical point of view \change{identifying} plasmonic effects is not trivial\change{, because they} require at least a full  2D Maxwell-consistent description\change{. Reduced} models like the unidirectional pulse propagation equation~\cite{PhysRevE.70.036604}, which are frequently used to describe plasma-based THz generation~\cite{PhysRevLett.105.053903,PhysRevLett.116.063902,Dey17}, are by construction not capable of capturing such resonant effects.    
	
	In this article, we consider the 2C-laser-induced plasma as a plasmonic structure, and investigate under which conditions such \change{a} perspective is important. In the context of nanoantennas (or metamaterials), e.g, for SH generation~\cite{PhysRevB.88.205125}, tailoring plasmonic resonances by tuning the shape of the plasmonic particle is a standard approach. Therefore, we follow a similar strategy and modify the usually prolate spheroidal plasma \change{shapes} to tri-axial \change{ellipsoids,} which can be achieved by using elliptically shaped laser beams. Depending on whether the laser polarization is oriented along the long beam axis~(\blue{along $y$}) or along the short beam axis~(\blue{along $x$}), plasmonic resonances are triggered or not~(see Fig.~\ref{fig:scheme}).
	\change{Because} nonlinear propagation effects are in both cases equally present, any difference between the THz emission spectra in \change{these} two cases is linked to plasmonic effects. 
	\change{Our experimental results detailed in Sec.~\ref{sec:exp} clearly evidence that THz pulses emitted 
		\blue{for laser polarization along the short beam axis} are shorter and have a much broader emission spectrum (see Fig.~\ref{fig:THz_experiment}).}
	%We demonstrate experimental results which reveal a significant difference: THz pulses are shorter and have a broader emission spectrum when the plasma is excited by the laser field in the direction with the short focal beam width and plasma width. 
	A simple analytical model \change{developed in Sec.~\ref{sec:theo}} allows us to link \change{this resonant broadening} to a leaky mode \change{of the ellipsoidal plasma}. \change{In particular, this model shows} that the resonance has a strong impact on the spectrum whenever electrons are excited along a direction where the plasma size is smaller than the plasma wavelength.
	Finally, \change{in Sec.~\ref{sec:mod}} direct three-dimensional (3D) Maxwell consistent simulations in tightly focused geometry confirm \change{our experimental and theoretical findings.}
	
	\section{Experimental results}
	\label{sec:exp}
	
	\begin{figure}[h]
		\begin{center}
			\includegraphics[width=1.0\columnwidth]{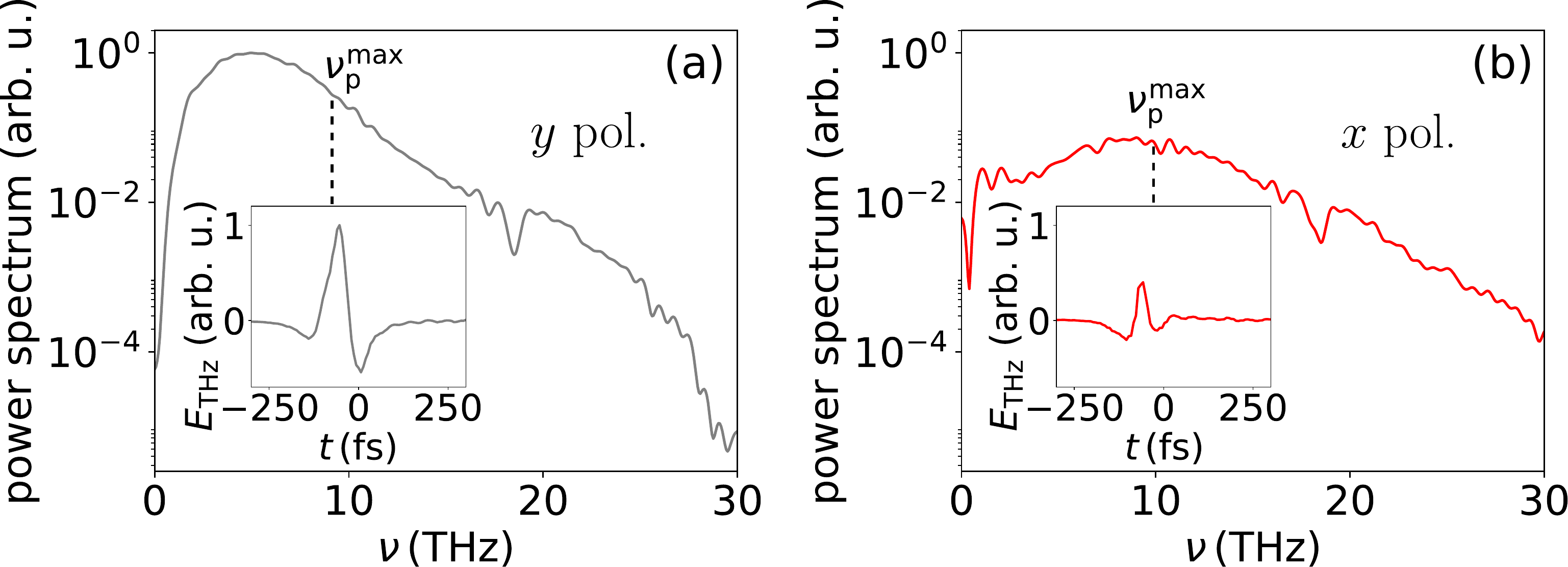}
		\end{center}
		\caption{\blue{Experimental THz spectra~(see text for details) with a $y$-polarized laser electric field (along the long axis of the elliptical beam) in (a) and an $x$-polarized laser electric field (along the short axis of the elliptical beam) in (b)}. Corresponding on-axis THz waveforms are shown as insets. The dashed lines specify the estimated maximum plasma frequency.}
		\label{fig:THz_experiment}
	\end{figure}
	
	\begin{figure}[h]
		\begin{center}
			\includegraphics[width=1.0\columnwidth]{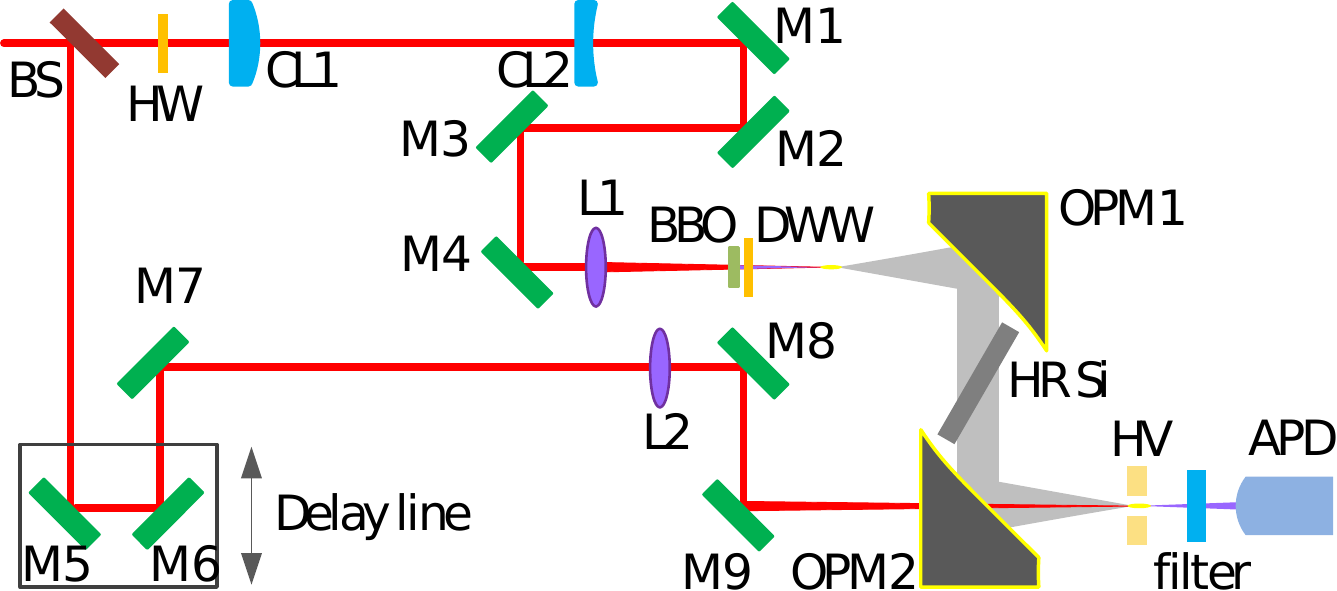}	
		\end{center}
		\caption{Sketch for 2C laser-induced air-plasma THz generation and detection system. BS: 800nm Beam splitter; HW: half wave plate; CL1: Plano-Convex cylindrical lens, focal length \change{$f = 100$~cm}; CL2: Plano-Concave cylindrical lens, \change{$f = -40$~cm}; M1-9: 45 degree incidence high reflective mirror; Lens L1: \change{$f = 30$~cm}; L2: \change{$f = 40$~cm}; DWW: Dual wavelength wave plate; OPM1 (OPM2): Off-axis parabolic mirrors, Reflected Focal Length $\mrm{RFL} = 4$~inch~(3~inch); HR Si: high-resistivity Silicon plate; $\mrm{RFL} = 3$~inch; HV: high voltage; APD: Si Avalanche Photodetector\change{; red line: laser beam; gray area: THz beam.}}
		\label{fig:experiment}
	\end{figure}
	
	\change{Experiments} are performed in a setup for two-color laser-induced air plasma THz generation and detection sketched in Fig.~\ref{fig:experiment}.
	The laser pulses originate from a \change{40-fs, 1-kHz} Ti:sapphire regenerative laser amplifier. The average laser power through the beam splitter (pump beam) is 230 mW; and the reflected beam (probe beam) has an average power of 200 mW.  The pump beam is vertically polarized by a half wave plate (800 nm) and sent into a 100-$\mu$m-thick BBO crystal through a \change{30-cm} focal lens for second harmonic generation (SHG). The polarization of the fundamental wavelength is then shifted back to horizontal after the SHG process, while the second harmonic wavelength is kept as horizontally polarized. Off-axis parabolic mirrors are used to collimate and focus the THz field. 
	% \rev{collecting all the forward emitted angular frequencies}
	A piece of high-resistivity silicon plate is employed to block the residual FH and SH laser beam. The probe beam, meanwhile, is focused by lens L2 and overlapped with the focus of the THz field through a central hole in OPM2. 
	The detection of the THz waveform is done by using the so-called air-biased coherent detection scheme~\cite{PhysRevLett.97.103903}\change{,} where in the presence of the THz field \change{and a high-voltage static field,} the four-wave-mixing-generated SH of the probe beam is measured. 
	%In the setup, we 
	An avalanche photodiode (APD) is employed as photodetector. %since it has more linear response compared to the normally adapted photomultiplier tube (PMT). 
	A boxcar integrator is then necessary to reshape the fast APD response (few ns) before it can be picked up by the lock-in amplifier. The part of the setup where THz waves are involved is covered by a plastic box and purged with dry nitrogen during the measurements \change{in order to avoid water absorption}. 
	\rev{This set-up enables the detection of signals up to 40~THz.} 
	A pair of cylindrical lenses is employed in the pump beam arm to elliptically shape the otherwise near-Gaussian pump beam profile. 
	\rev{The focal lengths of these lenses have been chosen in such a way that the ratio of the transverse extensions of the plasma is expected to be about 2.5.}
	We rotate both cylindrical lenses by $90^\circ$ to switch the pump electric field \blue{polarization with respect to the elliptically shaped beam and plasma profile}. The probe beam arm, on the other hand, is kept the same to ensure identical probe condition for THz generated by different pump beam modes.
	
	\begin{figure}
		\begin{center}
			\includegraphics[width=1.\columnwidth]{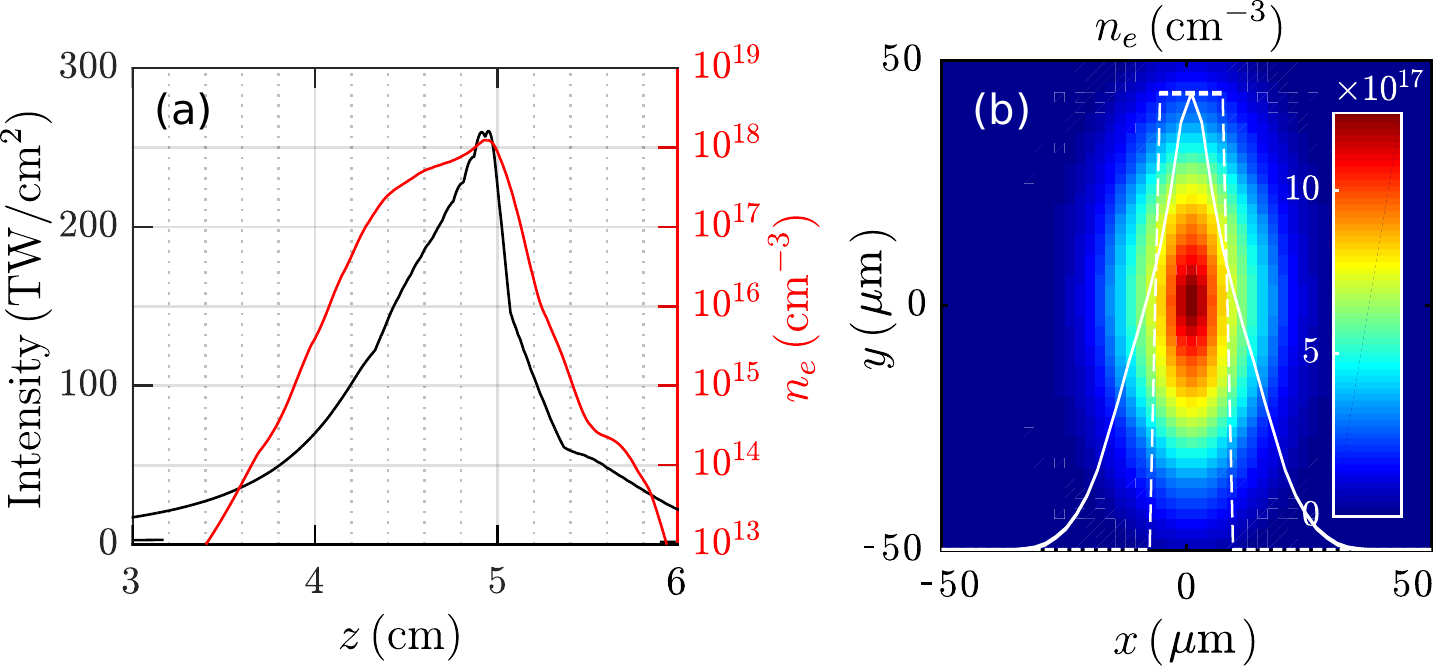}	
		\end{center}
		\caption{Results of the UPPE simulation: (a)~\rev{Maximum intensity~(black line) and electron density~(red line) in the transverse plane and in time along the propagation axis $z$}. (b)~Electron density profile \change{in the $(x,y)$ plane at the nonlinear focus $z\simeq5$~cm}. \rev{The line-out of the profile along $y=0$ is presented as solid white line. The approximation for the plasma slab model is sketched as dashed white line.}}
		\label{fig:UPPE}
	\end{figure}
	
	The experimental 2C elliptical beam creates an about \change{2.5-cm}-long plasma for any rotation angle of the cylindrical lenses. 
	We report no visible difference of the plasmas created \blue{in all the configurations}.
	To obtain more details about the \change{ellipsoidal} electron density profile under the present experimental conditions we performed simulations with the unidirectional propagation equation~(UPPE) approach~\cite{PhysRevE.70.036604}.
	As can be seen in Fig.~\ref{fig:UPPE}(a), the laser beam is focused after about 5~cm. 
	%Before the focus the elliptical beam is wider along $x$ than $y$ while in the focal plane the elliptical beam profile rotates and the beam is wider along $y$ than $x$~(not shown here). 
	In agreement with the experimental observations, the \change{230-$\mu$J laser pulse creates} a 2.5-cm-long plasma~[see Fig.~\ref{fig:UPPE}(a)]. The peak electron density grows to $n_\mrm{p}^\mrm{max}=1.2\times10^{18}\,\mrm{cm}^{-3}$ which corresponds to the maximum plasma frequency $\nu_\mrm{e}^\mrm{max}=\sqrt{q_\mrm{e}^2n_\mrm{e}^\mrm{max}/(m_\mrm{e}\epsilon_0)}/(2\pi)=9.9~\mrm{THz}$ and the minimum plasma wavelength $\lambda_\mrm{p}^\mrm{min}=30\,\mu$m. The transverse plasma shape \change{at focus} is strongly elliptical~[see Fig.~\ref{fig:UPPE}(b)]: The transverse FWHM extensions of the elliptical plasma were estimated to 20\,$\mu$m and 54\,$\mu$m \rev{giving a ratio of the transverse extensions of the plasma of 2.7 which is close to the expected value}. Thus, $\lambda_\mrm{p}^\mrm{min}$ is smaller than the plasma size in one direction and larger in the other direction which turn out to be the proper conditions to observe a significant difference in the THz emission spectra~(see Sec.~\ref{sec:theo}).
	
	%The experimental 2C elliptical beam creates an about 25-mm-long plasma for any rotation angle of the cylindrical lenses. By simulating the laser propagation by means of the unidirectional pulse propagation equation~\cite{PhysRevE.70.036604}, we could estimate the peak electron density to $n_\mrm{e}^\mrm{max}=1.2\cdot10^{18}$\,cm$^{-3}$ which translates into a maximum plasma frequency $\nu_\mrm{p}^\mrm{max}=\sqrt{(n_\mrm{e}^\mrm{max}q_\mrm{e}^2)/(m_\mrm{e}\epsilon_0)}/(2\pi)=9.9$\,THz and the minimum plasma wavelength $\lambda_\mrm{p}^\mrm{min}=30\,\mu$m. The transverse FWHM extensions of the elliptical plasma were estimated to 20\,$\mu$m and 54\,$\mu$m. Thus, $\lambda_\mrm{p}^\mrm{min}$ is smaller than the plasma size in one direction and larger in the other direction which turn out to be proper the conditions to observe a significant difference in the THz emission spectra.
	
	The detected \rev{angularly integrated} THz spectra and waveforms are presented in Fig.~\ref{fig:THz_experiment}. The \change{FWHM} THz pulse duration for \blue{$x$-polarization} is about 1.6 times shorter than for \blue{$y$-polarization}. A pyroelectric measurement \change{with a detection bandwidth limited to 12~THz} revealed a THz pulse energy ratio between \blue{$y$-polarization} and \blue{$x$-polarization} of 2.5. 
	The \blue{spectrum for $x$-polarization} is significantly broader than \blue{for $y$-polarization}. Moreover, the maximum THz emission for \blue{$x$-polarization} is found near the estimated maximum plasma frequency~(dashed line). \rev{Note that the spectral feature around 18 THz is the phonon band absorption of the high resistivity silicon plate which is used between the two off-axis parabolic mirrors to block the pump and second harmonic wavelengths.}
	
	%\change{Our results are a priori not limited to the gas-plasma configuration. Especially in solid- or the recently proposed liquid-plasmas~\cite{Dey17} where the electron density is large compared to gas-plasmas, plasmonic effects should play a crucial role. Besides THz generation also other frequency domains as the infrared might be concerned. Moreover, our conclusions for the IC mechanism can be easily extended to any nonlinear conversion process.}
	
	%\newpage
	
	\section{Theoretical explanation}
	\label{sec:theo}
	
	%In the following, we want to give a more detailed analysis of this 2D system and explain the difference
	In the following, we want to explain the difference in the THz \blue{emission} spectra \blue{between the $y$-polarization~(laser electric field polarized along the long beam axis) and $x$-polarization~(laser electric field polarized along the short beam axis)}. To this end we resort to a reduced model before \change{solving} the full \change{Maxwell-consistent} problem \change{in Sec.~\ref{sec:mod}}.
	
	For 2C-driving laser pulses and laser intensities of $10^{14}-10^{16}$~W/cm$^2$, the ionization current~(IC) mechanism is responsible for THz generation in gas-plasmas~\cite{Kim}. Thus, the THz-emitting current $\Jvec$ is given by
	\begin{align}
		\partial_t \Jvec + \nu_\mrm{ei}\Jvec &= \frac{q_\mrm{e}^2}{m_\mrm{e}} n_\mrm{e} \Evec\label{eq:current}\,\mbox{,}
	\end{align}
	with electron charge $q_\mrm{e}$, mass $m_\mrm{e}$ and electric field $\Evec$. The electron density $n_\mrm{e}$ is computed by means of rate equations employing a tunnel ionization rate \cite{Ammosov-1986-Tunnel,PhysRevA.64.013409}, and the electron-ion collision frequency $\nu_\mrm{ei}$ depends on the ion densities and the electron energy density~\cite{Huba2013}. \change{Equation~(\ref{eq:current}) coupled to Maxwell's equations} also appears as the lowest order of the expansion developed in \cite{PhysRevE.94.063202}, and fully comprises the IC mechanism~\cite{Thiele2017}.  
	
	Our reduced model requires three major approximations. First, we assume translational invariance in $y$ direction and solve a two-dimensional (2D) problem in $x$ and $z$ only. This means that the wider width of our elliptical beams is now infinite. 
	%In 2D qTE and qTM are then genuine TE and TM polarizations: the TE mode governs the $B_x,E_y,B_z$ components and the TM mode the $E_x,B_y,E_z$ components of the electromagnetic field. 
	The second major assumption concerns Eq. (1). The \change{main} nonlinearity appears due to the product of the time-dependent electron density $n_\mrm{e}$ and electric field $\Evec$. We split the electric field according to $\Evec=\tilde{\Evec}+\Evec_\mrm{L}$, where $\tilde{\Evec}$ is the field due to interaction of the laser with the plasma and $\Evec_\mrm{L}$ is the laser field defined by its \change{2D} propagation in vacuum. For a qualitative understanding of the phenomenon it is sufficient to assume the paraxial solution for the laser electric field \change{$\Evec_{\rm L}$}.
	Then, the right-hand-side of Eq.~(\ref{eq:current}) %inside the slab 
	reads $\frac{q_\mrm{e}^2}{m_\mrm{e}} n_\mrm{e} \tilde{\Evec} + \frac{q_\mrm{e}^2}{m_\mrm{e}} n_\mrm{e} \Evec_\mrm{L}$. 
	\change{An additional simplification, which allows further analytical treatment,} is now to replace in the first term the time-dependent electron density \change{$n_\mrm{e}(t)$} by a time-invariant electron density $n_0$,
	but keep \change{$n_\mrm{e}(t)$} in the second term (THz current source) accounting for ionization by the laser. By doing so, we consider \change{a plasma with electron density $n_0$} to be excited by the current source $\bm\iota=\frac{q_\mrm{e}^2}{m_\mrm{e}}n_\mrm{e}\Evec_\mrm{L}$. 
	%We choose the time-invariant electron density as the electron density after ionization $n_0=n_\mrm{e}(t=\infty)$. The coupling to the Maxwell's equations results in a linear system of equations~(for $\nu_\mrm{ei}=\mrm{const}$). This approach permits to investigate the THz emitting laser-induced gas-plasma as a plasmonic particle with the electron density profile $n_0(\rvec)$ where $\rvec=(x,y,z)^\mrm{T}$ is the spatial coordinate. To this end, we can use the well developed "plasmonic toolbox" and analyze the resonances which are supported by the gas-plasma and their emission properties. To analyze the behavior of the elliptically shaped plasma analytically, we simplify the plasma shape assuming a 
	\change{The third assumption is finally to consider a} plasma slab of thickness $d$ as sketched in Fig.~\ref{fig:slab}(a) with translational invariance in $y$ and $z$. %\cite{Kostin:10,Kostin15,doi:10.1063/1.4953098}
	%In short, we consider a system as sketched in Fig.~\ref{fig:slab}(a): A plasma slab with thickness $d$ along the $x$-direction characterized by a time-invariant electron density $n_0$ and collision frequency $\nu_\mrm{ei}$. Both quantities are assumed to be translational invariant in $y$ and $z$. 
	Above and below the slab we assume a semi-infinite vacuum. 
	\change{For the time-invariant electron density in the slab we choose the overall peak density $n_{\rm e}^{\rm max}$, and we can write}
	\begin{equation}
	\change{n_0 = n_{\rm e}^{\rm max} \theta(x-d/2)\theta(d/2-x)\,\mrm{,}}
	\end{equation}
	\change{where $\theta$ denotes the usual step or Heaviside function.}
	
	\begin{figure}[t]
		\centering
		\includegraphics[width=1.0\columnwidth]{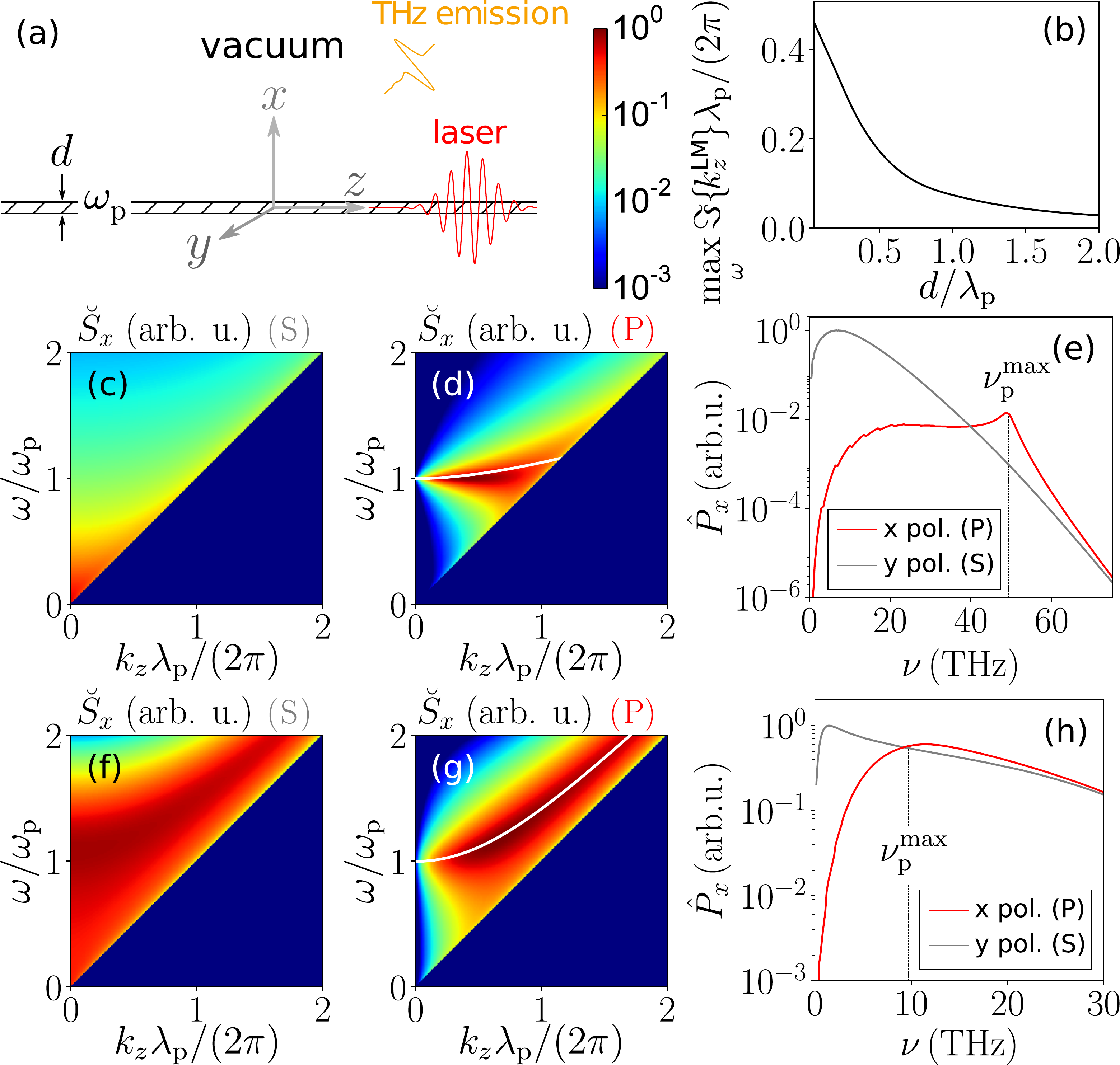}
		\caption{(a)~Illustration of the plasma slab model. (b)~Imaginary part \rev{$\Im\{k_z^{\rm LM}\}$} of the leaky mode propagation constant versus the slab thickness $d$ \blue{in P polarization}. Transverse Poynting flux for a $\delta$ excitation (see text for details) \blue{for S~(c) and P~(d) polarization}. In (d), the real part \rev{$\Re\{k_z^{\rm LM}\}$} of the leaky mode propagation constant is indicated by a white line. (e)~Far-field spectra for \blue{S and P polarization} predicted by the plasma slab model employing source terms obtained from 2D paraxial laser field propagation, reproducing well 2D and 3D simulation results from Fig.~\ref{fig:spectra_elipt}. The slab thickness is $d=0.4\,\mu$m, the plasma frequency $\omega_\mrm{p}/(2\pi)=\nu_\mrm{p}=49.15$~THz. (f-h)~Same as (c-e), but for $d=15\,\mu$m and $\nu_\mrm{p}=10$~THz, reproducing well the experimental results from Fig.~\ref{fig:THz_experiment}.}
		\label{fig:slab}
	\end{figure}
	
	Let us first discuss the response of such \change{a} plasma slab, and solve the reflection transmission problem for an incident plane wave in the $x,z$ plane.
	\blue{
		In the context of a reflection transmission problem the $y$-polarized electric field configuration is often called S polarization, and the $x$-polarized configuration is referred to as P polarization: the S polarization mode governs the $B_x,E_y,B_z$ components and the P polarization mode the $E_x,B_y,E_z$ components of the electromagnetic field.}
	%In the waveguide context, the S-polarization is sometimes identified with transverse magnetic~(TM) waves and the P-polarization is sometimes identified with transverse magnetic~(TE) waves.
	% In 2D qTE and qTM are then genuine TE and TM polarizations: the TE mode governs the $B_x,E_y,B_z$ components and the TM mode the $E_x,B_y,E_z$ components of the electromagnetic field. 
	
	\change{For sake of simplicity, we neglect losses in the slab and set $\nu_{ei}=0$.} The reflectivity for \blue{S and P polarization} reads\change{~(see~supplementary materials \cite{suppl} for details)}
	\begin{equation}
	\rho_{\rm S/P} = \left|\frac{{\Lambda^{\rm v}}^2-{\Lambda^{\rm p}}^2\alpha_{\rm S/P}^2}
	{{\Lambda^{\rm v}}^2+{\Lambda^{\rm p}}^2\alpha_{\rm S/P}^2
		+\frac{2{\Lambda^{\rm v}} {\Lambda^{\rm p}}\alpha_{\rm S/P}}{\tanh\!\left({\Lambda^{\rm p}}d\right)}}\right|^2, \label{eq:rho_slab} 
	\end{equation}
	with $\Lambda^{\mrm{v}}=\sqrt{k_z^2 - \omega^2/c^2}$, $\Lambda^{\mrm{p}}=\sqrt{k_z^2 - \epsilon^{\mrm{p}}\omega^2/c^2}$, $\alpha_{\rm S}=1$, 
	$\alpha_{\rm P}=1/\epsilon^{\mrm{p}}$, and plasma dispersion $\epsilon^\mrm{p}=1-\omega_\mrm{p}^2/\omega^2$. Singularities of the reflectivity correspond to guided modes, and it is known that for frequencies below the plasma frequency $\omega_\mrm{p}$ two surface plasmon polaritons (SPPs) exist for \blue{P} polarization~\cite{Berini:09}. 
	%These modes are usually termed long-range and short-range SPP, and 
	Their real propagation constants \rev{$k_z^{\rm SPP}$} are zeros of the denominator of Eq.~(\ref{eq:rho_slab}). For \blue{S polarization}, no such guided modes exist. What turns out to be very relevant for the following analysis is that for \blue{P polarization} the plasma slab also features a \emph{leaky mode} slightly above the plasma frequency $\omega_\mrm{p}$. A leaky mode solution has a complex propagation constant \rev{$k_z^{\rm LM}$}. It features a Poynting flux away from the guiding structure, in our case the plasma slab, and its amplitude decreases exponentially upon propagation in $z$ direction, which is why it is termed \emph{leaky}. 
	
	%It also has the ``uncomfortable property'' that it grows exponentially as $x \rightarrow \pm \infty$.
	The leaky mode is naturally coupled to the radiative field, and can represent a solution of the driven system with the source~\cite{Marcuvitz56,Oliner1959,Hu2009,Kostin15}. 
	In order to find the leaky mode for the plasma slab, one has to substitute $\Lambda^{\mrm{v}}\rightarrow-\Lambda^{\mrm{v}}$ in Eq.~(\ref{eq:rho_slab}), that is, switch the Poynting flux in the vacuum away from the slab, and look for complex zeros of the denominator:
	\begin{equation}\label{eq:leaky_TM}
	\tanh\!\left({\Lambda^{\rm p}}d\right) = \frac{2{\Lambda^{\rm v}} {\Lambda^{\rm p}}\epsilon^\mrm{p}}{{\Lambda^{\rm v}}^2{\epsilon^\mrm{p}}^2+{\Lambda^{\rm p}}^2}.
	\end{equation}
	For given frequency $\omega$ and thickness $d$, the imaginary part \rev{$\Im\{k_z^{\rm LM}\}$} of the complex solution to Eq.~(\ref{eq:leaky_TM}) determines how strong radiation \emph{leaks} from the plasma slab to the surrounding vacuum. In Fig.~\ref{fig:slab}(b), the maximum of \rev{$\Im\{k_z^{\rm LM}\}$} versus the slab thickness $d$ is shown, clearly indicating that only for thin plasma slabs~($d<\lambda_\mrm{p}$) the leaky mode should play a role in the emission spectrum.
	
	%Next, we want to link the plasma slab results with the laser driven gas plasma, that is, with Eq.~(\ref{eq:current}) coupled to Maxwell's equations.
	%For sake of simplicity, we ignore collisions $(\nu_\mathrm{ei}=0)$.
	
	Let us now examine the response of the plasma slab to a $\delta$-excitation $\bm\iota\propto \delta(z)\delta(t)$ for $|x|<d/2$ along $y$~(\blue{S}) and $x$~(\blue{P}). Such excitation is constant in the $\omega,k_z$ Fourier domain. 
	To collect the emission radiating from the slab, we consider the $x$-component of the spectral Poynting flux $\breve{S}_x\propto\Re\{\breve{\Evec}\times\breve{\Bvec}^\star\}_x$ \change{through two planes normal to the $x$ axis situated above and below the slab.} 
	In Figs.~\ref{fig:slab}(c,d), $\breve{S}_x$ is visualized for a thin slab $d=0.066\lambda_p$. The resonant emission following the real part \rev{$\Re\{k_z^{\rm LM}\}$} of the leaky mode for \blue{$x$} polarization (white line) is evident. For the larger slab with $d=0.5\lambda_p$ shown in Figs.~\ref{fig:slab}(f,g), the resonance associated with the leaky mode moves away from $\omega_{\rm p}$%, and \change{becomes broader}.
	
\begin{figure}[h]
		\centering
		\includegraphics[width=0.85\columnwidth]{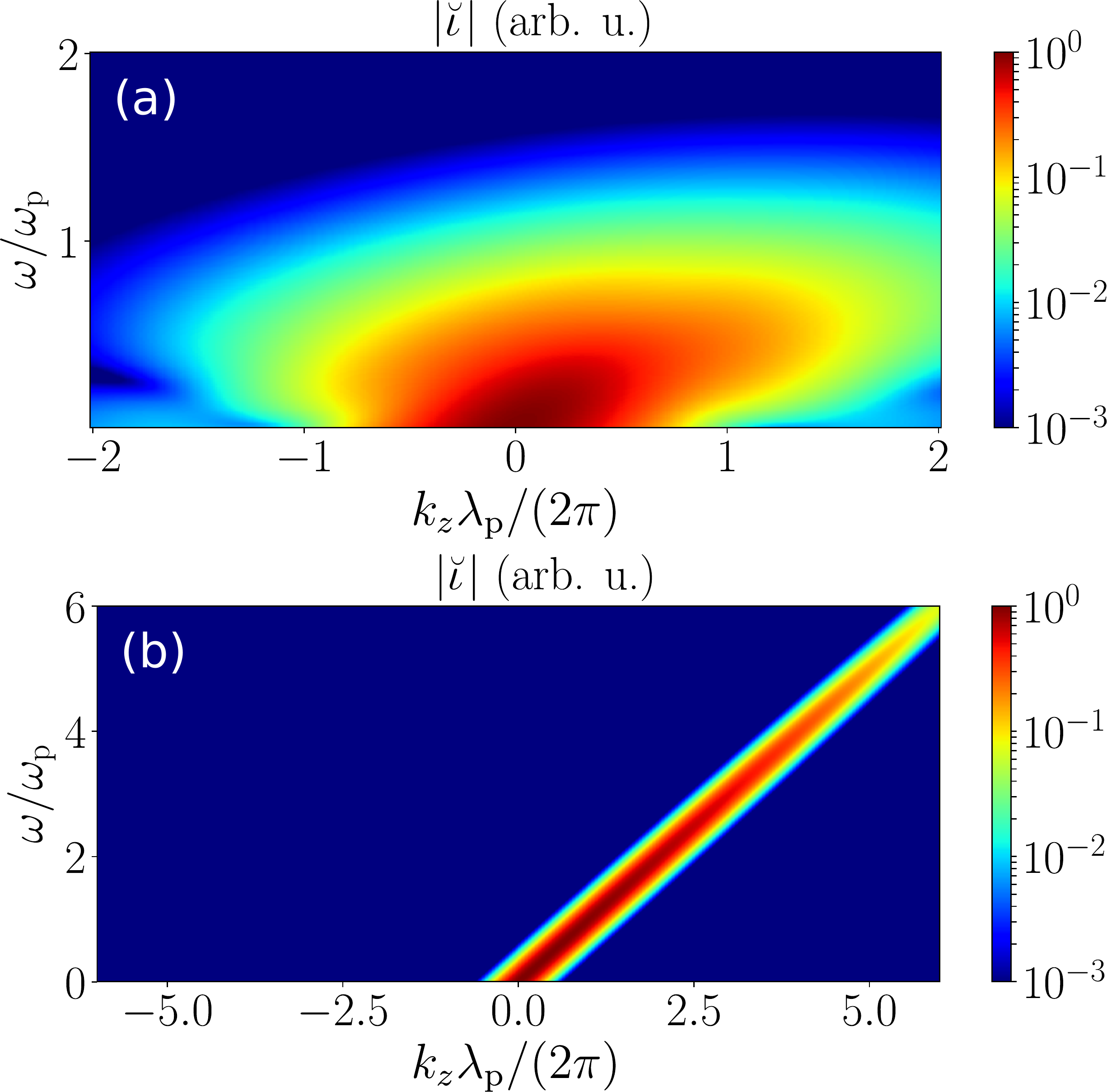}
		\caption{Excitation source terms which are used to model the THz emission spectra \change{presented in Fig.~\ref{fig:slab}(e,h): (a)~strongly focused configuration leading to a thin plasma, (b)~weakly focused configuration leading to a thicker plasma}.}
		\label{fig:iota}
	\end{figure}

	In order to make a more quantitative comparison with previous experiments and upcoming simulation results, we now approximate the source term $\bm\iota$ by 2D paraxial laser field propagation and the corresponding time and $z$ dependent electron density on the optical axis. The electric field on the optical axis in quasi-monochromatic 2D paraxial approximation reads
	\begin{align}
		\Evec_\mrm{L}(z,t) \approx &\Re \left\{\frac{E_{\omega} \, e^{-\frac{\left(t - z/c\right)^2}{t_0^2}-\rmi\omega_\mrm{L}\left(t - z/c\right) }}{\sqrt{1+\rmi\frac{z}{z_\mrm{R}(\omega_\mrm{L})}}} \right\}\evec_\mrm{L} \nonumber\\
		& +\Re \left\{\frac{E_{2\omega} \, e^{-\frac{\left(t - z/c\right)^2}{t_0^2}-\rmi2\omega_\mrm{L}\left(t - z/c\right) -\rmi\phi}}{\sqrt{1+\rmi\frac{z}{z_\mrm{R}(2\omega_\mrm{L})}}} \right\}\evec_\mrm{L}\,\mbox{,}
	\end{align}
	where $E_{\omega}$, $E_{2\omega}$ are the FH or SH electric field amplitudes, $t_0$ is the \change{$1/e$} pulse duration, $\omega_\mrm{L}=2\pi c/\lambda_\mrm{L}$ the FH laser frequency with wavelength $\lambda_\mrm{L}$, $\phi$ the relative phase angle \change{between FH and SH}, the unit vector $\evec_\mrm{L}$ defines the (linear) laser polarization direction, $z_\mrm{R}(\omega)=w_{0,x}^2\omega/2c$ is the Rayleigh length and the symbol $\Re$ denotes the real part of a complex quantity. Then the model current source in the slab reads
	\begin{equation}
	\bm\iota = \frac{q_\mrm{e}^2 n_\mrm{e}[\Evec_\mrm{L}]}{m_\mrm{e}}\Evec_\mrm{L}\,\mrm{.}
	\end{equation}
	As before, we assume $\bm\iota$ to be invariant along $x$ inside the slab and zero outside for simplicity. 
	The current source for a tightly focused configuration is displayed in Fig.~\ref{fig:iota}(a). In this case, the excitation leads to slightly forward oriented terahertz emission. The excitation spectrum for a weakly focused beam is presented in Fig.~\ref{fig:iota}(b). Here, the excitations happens close to the \change{vacuum dispersion relation} line $k_z(\omega)=\omega/c$. As the result, THz waves are emitted rather in a forward cone.
	
	Corresponding angularly integrated far-field power spectra $\hat{P}_x$ for the 2D plasma slab model are shown in Figs.~\ref{fig:slab}(e,h). The resulting THz emission spectra \rev{in Fig.~\ref{fig:slab}(h)} reproduce well the qualitative behavior of the experimental spectra in Fig.~\ref{fig:THz_experiment}: In contrast to \blue{$y$-polarization (S)}, the emission spectrum for \blue{$x$-polarization (P)} is broadened \change{around} the maximum plasma frequency $\nu_\mrm{p}^\mrm{max}$. For the \change{thinner} slab and \change{higher} plasma frequency in Fig.~\ref{fig:slab}(e), which rather \change{mimics} the situation of the \change{upcoming simulations}, the resonance peak \blue{in the spectrum for $x$-polarization} is \change{stronger and less broad}, as expected from the \change {behavior} of the leaky mode.
	
	Before continuing with full Maxwell consistent simulation results, let us \change{give an intuitive} explanation for the difference in the \blue{THz spectra in the low frequency range well below $\omega_{\rm p}$.} 
	When the plasma current is excited along the \change{invariant $y$} direction, \change{no} charge separation \change{occurs} due to the displacement of the electrons with respect to the ions. Thus, \change{there is no} restoring force and \change{low frequency currents} can persist \change{and emit radiation}. In contrast, when the plasma current is excited along the $x$ direction \change{with} strong plasma gradients, a significant charge separation force pulls the electrons back to the ions 
	\change{which prevents the generation of low-frequency currents. In 3D this scenario also holds, the charge separation effects are more pronounced for \blue{laser polarization along the short beam axis} which explains the smaller THz signal in the lower frequency range in experiments and simulations \blue{for this case}~[see Figs.~\ref{fig:THz_experiment},~\ref{fig:spectra_elipt}].}
	%resulting in higher frequency components in the current and the emitted radiation. 
	
	\section{Maxwell consistent modeling}
	\label{sec:mod}
	
	In the following, we consider the results of full Maxwell consistent simulations obtained with the code ARCTIC~\cite{Thiele2017}. Here, the electron density is space- and time-dependent and the electric field is treated self-consistently by coupling the \change{complete} current Eq.~(\ref{eq:current}) to the Maxwell's equations. \change{Maxwell-consistent} simulations are computationally expensive\change{, and modeling} the experimental configuration with cm-long gas-plasmas \change{is beyond the possibilities of current supercomputer clusters.} In contrast, when focusing the laser pulse strongly, the gas-plasma can be only tens-of-$\mu$m long and less than one $\mu$m thick~\cite{Thiele2017}. \change{As can be anticipated from Fig. 5,} the small transverse size is favorable to excite a leaky mode \blue{for the $x$-polarized laser} case.  %Due to the small plasma length, the plasma defocussing of the laser beam is weak which leads to laser propagation almost like in vacuum. Thus, intensities high enough to fully singly ionize a gas at ambient pressure can be reached. 
	
	We define the driving laser pulse by \change{prescribing} its transverse vacuum electric field at focus \change{as}
	\begin{equation}
	\begin{split}
	\Evec_{\mrm{L},\perp}& (\rvec_\perp, z=0, t) = \exp\!\left(-\frac{x^2}{w_{0,x}^2}-\frac{y^2}{w_{0,y}^2}-\frac{t^2}{t_0^2}\right)\\
	&\times\bigg[E_{\omega}\cos\!\left(\omega_\mrm{L}t\right)+E_{2\omega}\cos\!\left(2\omega_\mrm{L}t+\phi\right)\bigg]\evec_\mrm{L}\mbox{,}
	\label{eq:laser_elipt}
	\end{split}
	\end{equation}
	where $w_{0,x}$ is the short and $w_{0,y}$ is the long vacuum focal beam width. The laser pulse propagates in the positive $z$ direction, and the origin of the coordinate system is at the vacuum focal point. By defining the tightly focused laser pulse at vacuum focus we follow the algorithm described in~\cite{Thiele20161110}.
	
	\begin{figure}[t]
		\begin{center}
			\includegraphics[width=1.0\columnwidth]{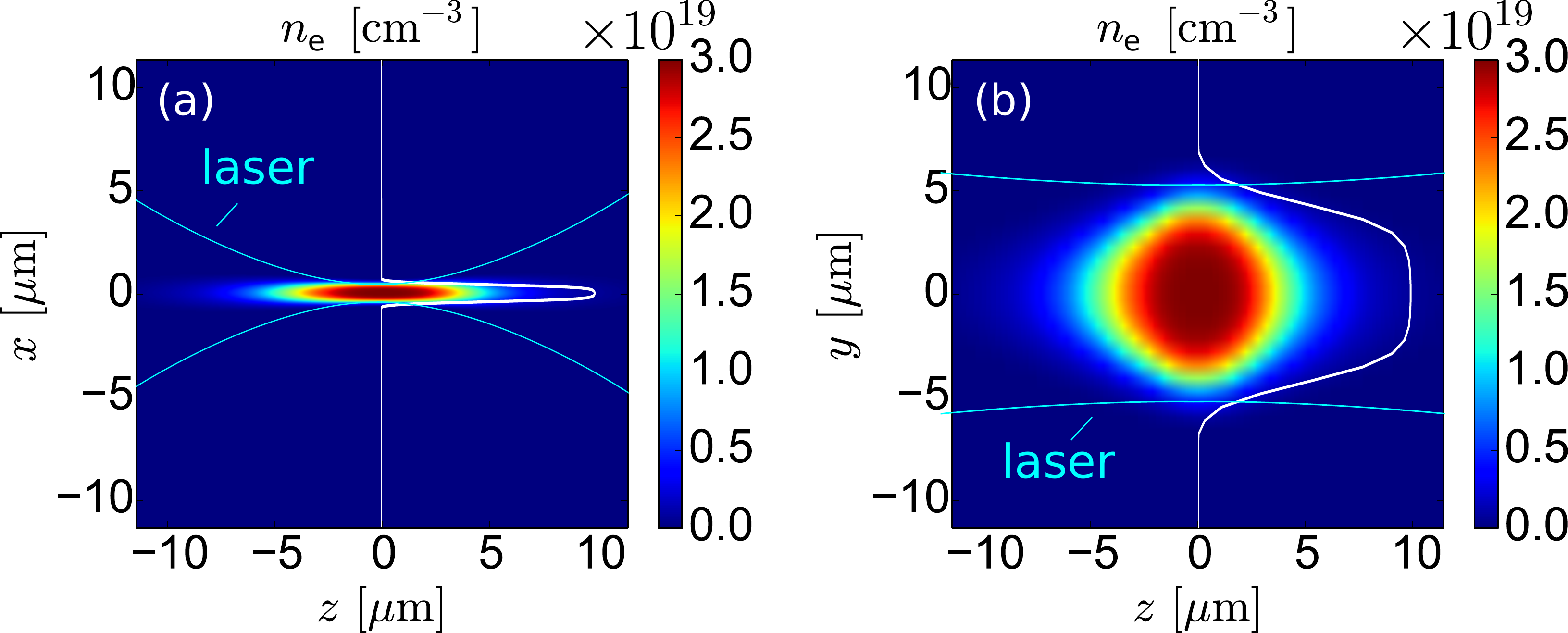}	
		\end{center}
		\caption{Electron density in (a)~$zx$ and (b)~$yz$ planes after ionization of an Argon gas with initial neutral density ${n_\mrm{a}=3\times10^{19}}$~cm$^{-3}$~($\approx1$~bar) by a 2C elliptically shaped laser pulse~($E_{\omega}=40$~GV/m, $E_{2\omega}=20$~GV/m, $t_0=50$~fs, $w_{0,x}=\lambda_\mathrm{FH}=0.8~\mu$m, $w_{0,y}=8~\mu$m). The respective electron density profiles at focus ($z=0$) are visualized by the bright white lines. The waist of the focused laser is tracked by the light blue lines.}
		\label{fig:Ey_n0_qTE}
	\end{figure}
	%The laser defined by Eq.~(\ref{eq:laser_elipt}) has the short beam width $w_{0,x}$ and the long beam width $w_{0,y}$~(see Fig.~\ref{fig:Ey_n0_qTE}) resulting in a spatially elliptical Gaussian beam profile. 
	The laser pulse parameters are chosen such that in argon with initial neutral density ${n_\mrm{a}=3\times10^{19}}$~cm$^{-3}$~($\approx1$~bar) a fully singly ionized ellipsoidal plasma is created~(see Supplemental Material~\cite{suppl} for details). 
	The peak electron density $n_\mrm{e}^\mrm{max}=n_\mrm{a}$ translates into a maximum plasma frequency $\nu_\mrm{p}^\mrm{max}=\sqrt{(n_\mrm{a}q_\mrm{e}^2)(m_\mrm{e}\epsilon_0)}/(2\pi)\approx50$~THz and a minimum plasma wavelength ${\lambda_\mathrm{p}^\mrm{min}=c/\nu_\mrm{p}^\mrm{max}\approx6\,\mu}$m. 
	As visualized in Fig.~\ref{fig:Ey_n0_qTE}, the transverse plasma profile is strongly elliptical, that is, along $x$ direction the plasma size is less than 1~$\mu$m, whereas along $y$ direction the plasma is approximately 10~$\mu$m wide.
	\blue{By setting the linear laser polarization $\evec_\mrm{L}=\evec_y$ or $\evec_\mrm{L}=\evec_x$}, we thus excite a THz emitting current $\Jvec$ along $y$ direction where the plasma profile is wide or along $x$ direction where the plasma profile is narrow.%, respectively. 
	% !!!
	% Note that the electron density profile itself does not depend on the laser polarization direction, because due to the short interaction length the laser propagates without any noticeable deformation.
	
	Considering the forward emitted THz radiation in Fig.~\ref{fig:spectra_elipt}, we find strong single-cycle pulses reaching field amplitudes of 10~kV/cm. 
	Most importantly, the THz pulse obtained for \blue{$x$-polarization}~(b) is two times shorter than for \blue{$y$-polarization}~(a), as a direct consequence of the THz emission spectra that are dramatically different.
	3D angularly integrated THz far-field spectra in Fig.~\ref{fig:spectra_elipt} reveal that \blue{for $x$-polarization} the THz spectrum is broadened up to about 50~THz, the maximum plasma frequency $\nu_\mrm{p}^\mrm{max}$, while \blue{for $y$-polarization} no such broadening is found. The total THz pulse energy~($\nu<70$~THz) \blue{for $y$-polarization} is 4.8 times larger than \blue{for $x$-polarization}. 
	
	%% Separation TE/TM:
	\begin{figure}[h]
		\centering
		\includegraphics[width=1.0\columnwidth]{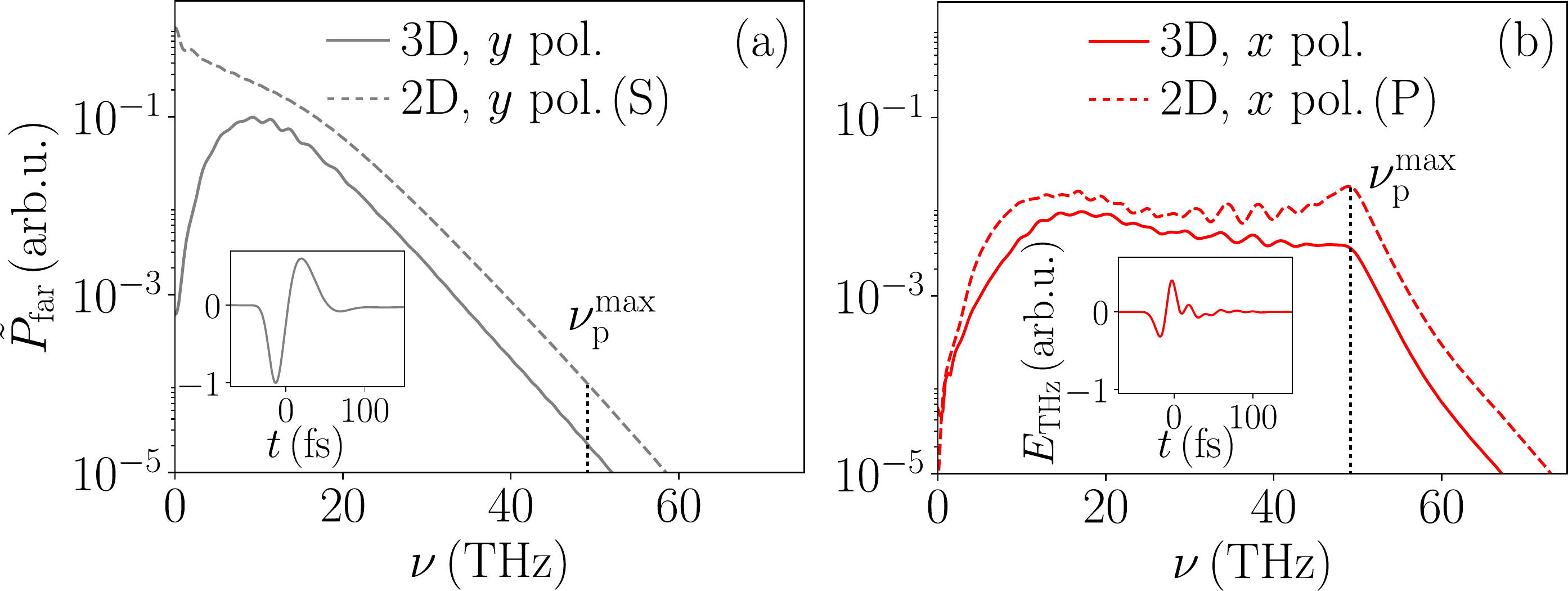}
		\caption{Angularly integrated far-field spectra for the elliptical beams from Fig.~\ref{fig:Ey_n0_qTE} (solid lines) and corresponding results from 2D simulations assuming translational invariance in $y$ (dashed lines). The insets show the forward emitted THz pulses recorded at $z=12.7\,\mu$m behind the plasma.}
		% are presented as white lines demonstrating a significantly shorter pulse duration for qTM polarization, which can be attributed to triggering a plasmonic resonance~($E_{\omega}=40$~GV/m, $E_{2\omega}=20$~GV/m, $t_0=50$~fs, $w_{0,x}=\lambda_\mathrm{FH}=0.8~\mu$m, $w_{0,y}=8~\mu$m)
		\label{fig:spectra_elipt}
	\end{figure}
	
	%Before performing a more detailed theoretical analysis, let us give a simple explanation for the difference in the THz spectra for qTE and qTM polarization: When the plasma current is excited along the direction of weak plasma gradients, the charge separation created due to the displacement of the electrons with respect to the ions is relatively small. Thus, the restoring force is small and the electron current can persist for a long time which leads to a quasi-DC current resulting in the emission of rather low frequencies. In contrast, when the plasma current is excited along the direction of strong plasma gradients, a significant charge separation force pulls the electrons back to the ions resulting in higher frequency components in the current and the emitted radiation. 
	
	The dashed lines in Fig.~\ref{fig:spectra_elipt} show the results of corresponding 2D simulations, i.e., $w_{0,y}\rightarrow\infty$. In this limit, we find a similar behavior: no broadening if the laser electric field is oriented in the now translationally invariant $y$ direction \blue{(involving $B_{x}$, $E_{y}$, $B_{z}$, S polarization)}, and broadening up to $\nu_\mrm{p}^\mrm{max}$ if the laser electric field points in $x$ direction \blue{(involving $E_{x}$, $B_{y}$, $E_{z}$, P polarization)} where the plasma profile is narrow.
	This observation additionally justifies our restriction to 2D for our theoretical model in the previous section.
	%Moreover, the TE polarization state governs the field components $B_{x}$, $E_{y}$, $B_{z}$, and the TM polarization state governs the field components $E_{x}$, $B_{y}$, $E_{z}$. 
	\change{Note that any} other polarization state in 2D geometry can be written as the superposition of these two cases, provided that the plasma density remains unchanged. We checked that this property also holds for 3D elliptical beams~(see~supplementary materials \cite{suppl} for details). This possibility of superposing \blue{$x$- and $y$-polarizations} implies that the THz emission spectrum can be tuned by rotating the linear polarization of the incoming laser pulse. 
	
	\section{Conclusion}
	In summary, we have shown that plasmonic effects can significantly broaden the terahertz emission spectrum from fs-laser-induced gas-plasmas when the corresponding \change{plasma} wavelengths are larger than the transverse plasma size.
	%Then, a sufficient condition to obtain the broadening is that the THz-emitting current is oriented along the direction of the small plasma dimension. 
	In the framework of a simplified model considering a plasma slab, we identified the plasmonic resonance responsible for the observed spectral features as a leaky mode.
	Our model seems to be predictive in microplasma configuration as well as for larger plasmas, as they are employed in standard two-color setups for THz generation.
	We propose an efficient THz-generation scheme to access this effect by two-color elliptically-shaped laser-beam-induced gas-plasmas via the ionization current mechanism and demonstrate its experimental feasibility. 
	\change{By turning} the polarization of the excitation, when adjusting the linear laser polarization, we can switch THz spectral broadening due to plasmonic resonances on or off and thus control the THz emission spectrum. 
	%We believe that this plasmonic view on gas-plasma-based THz generation paves a way towards resonant control of THz generation.}
	%We believe that our results, which are a priori not limited to the gas-plasma configuration, will trigger further experimental and theoretical efforts in this direction.
	We believe that considering the gas plasma as a THz plasmonic particle paves the way towards active resonant control of THz generation, and will trigger further experimental and theoretical efforts in this direction.
	
	\section*{Acknowledgments}
	Numerical simulations were performed using computing resources at
	M\'esocentre de Calcul Intensif Aquitaine (MCIA), Grand {\'E}quipement
	National pour le Calcul Intensif (GENCI, Grants No.~A0020507594 and No.~A0010506129) and Chalmers Centre for Computational Science and Engineering (C3SE) provided by the Swedish National Infrastructure for Computing (SNIC, Grants No.~SNIC2018-4-12, SNIC-2017-1-484 and SNIC2018-4-38). This study was supported by ANR
	(Projet ALTESSE). 
	\blue{I.T. and E.S. acknowledge support by the Knut and Alice Wallenberg Foundation.}
	S.S.\ acknowledges support by the Qatar National Research Fund through the National Priorities Research Program (Grant No.\ NPRP 8-246-1-060).
	
	%%%%%%%%%%%%%%%%%%%%%%% References %%%%%%%%%%%%%%%%%%%%%%%%%
	
	\begin{appendix}
		\section{Definitions: Fourier Transforms and Far-field Spectra}
\label{app:FT}

In the following, we define the temporal Fourier transform $\hat{f}(\rvec, \omega)$ of a function $f(\rvec, t)$ by
\begin{align}
\hat{f}(\rvec, \omega) &= \frac{1}{2\pi}\int f(\rvec, t)e^{\rmi\omega t}\,dt\,\mrm{,}\label{eq:IFT}\\
f(\rvec, t) &= \int \hat{f}(\rvec, \omega)e^{-\rmi\omega t}\,d\omega \label{eq:FT}\,\mbox{.}
\end{align}
Furthermore, we define the longitudinal spatial Fourier transform $\breve{f}(\rvec_\perp, k_z, \omega)$ of a function \change{$\hat{f}(\rvec_\perp,z, \omega)$} by
\begin{align}
\breve{f}(\rvec_\perp, k_z, \omega) &= \frac{1}{2\pi}\int \hat{f}(\rvec_\perp,z, \omega)e^{-\rmi k_z z}\,dz\,\mrm{,}\label{eq:FT_2kz}\\
\hat{f}(\rvec_\perp, z, \omega) &= \int \breve{f}(\rvec_\perp,k_z, \omega)e^{\rmi k_z z}\,d\change{k_z}\label{eq:space_fromkz}\,\mbox{.}
\end{align}
Note the difference in sign of the exponent for temporal and spatial transforms, which is common practice in the optical context. 

We introduce the spectral poynting fluxes
\begin{equation}
\hat{\Svec}(\rvec_\perp, z, \omega)=2/\mu_0\Re\{\hat{\Evec}(\rvec_\perp, z, \omega)\times\hat{\Bvec}^\star(\rvec_\perp, z, \omega)\}\,\mrm{,}
\end{equation}
and
\begin{equation}
\breve{\Svec}(\rvec_\perp, k_z, \omega)=2/\mu_0\Re\{\breve{\Evec}(\rvec_\perp, k_z, \omega)\times\breve{\Bvec}^\star(\rvec_\perp, k_z, \omega)\}\,\mrm{,}\label{eq:S}
\end{equation}
\change{where $\mu_0$ is the magnetic permeability and $\Re$ denotes the real part.}
The first expression is used to compute the angularly integrated far-field power spectrum in Maxwell-consistent 2D/3D simulations by integration over a closed surface around the plasma. \change{To} compute the angularly integrated far-field spectrum $\hat{P}_x$ in \change{Sec.~3 of the main article}, we integrate $\breve{S}_x$ along $k_z$.

\section{Modeling the ionization current mechanism for microplasmas}
\label{app:num}
The THz generation by two-color laser pulses considered here is driven by the so-called ionization current (IC) mechanism~\cite{Kim}. A comprehensive model based on the fluid equations for electrons describing THz emission has been derived in~\cite{PhysRevE.94.063202}. In this framework, the IC mechanism naturally appears at the lowest order of a multiple-scale expansion. Besides the IC mechanism, this model is also able to treat THz generation driven by ponderomotive forces and others. They appear at higher orders of the multiple-scale expansion and are not considered here. In the following, we briefly summarize the equations describing the IC mechanism.

The electromagnetic fields $\Evec$ and $\Bvec$ are governed by Maxwell's equations in vacuum
\begin{align}
\nabla\times\Evec &= - \partial_t \Bvec\,\mrm{,}
\label{eq:Far_1}\\
\nabla\times\Bvec &= \frac{1}{c^2}\partial_t \Evec + \mu_0 \Jvec + \mu_0 \Jvec_\mrm{loss}\,\mbox{.}
\label{eq:Amp_1}
\end{align}
The plasma and electromagnetic fields are coupled via the conductive current density $\Jvec$ governed by
\begin{align}
\partial_t \Jvec + \nu_\mrm{ei}\Jvec &= \frac{q_\mrm{e}^2}{m_\mrm{e}} n_\mrm{e} \Evec\,\mbox{,}\label{eq:cont}
\end{align}
where electron-ion collisions lead to a damping of the current. The collision frequency is determined by~\cite{Huba2013}
\begin{equation}
\nu_\mrm{ei}[\mrm{s}^{-1}] = \frac{3.9 \times 10^{-6}\sum\limits_Z Z^2 n_\mrm{ion}^{(Z)}[\mrm{cm}^{-3}]\lambda_\mrm{ei}}{E_{\mrm{elec}}[\mrm{eV}]^{3/2}}  \,\mbox{,}
\label{eq:nu_ei_NRL_2}
\end{equation}
where $\lambda_\mrm{ei}$ is the Coulomb logarithm \change{and $E_{\rm elec}$ denotes the thermal and kinetic electron energy}. The value $\lambda_\mrm{ei}=3.5$ turned out to match the results obtained by more sophisticated calculations with Particle-In-Cell (PIC) codes in~\cite{PhysRevE.94.063202}. The densities of $Z$ times charged ions are determined by a set of rate equations
\begin{equation}
\begin{split}
\partial_t n_\mrm{ion}^{(Z)} & = W^{(Z)} n_\mrm{ion}^{(Z-1)} - W^{(Z+1)} n_\mrm{ion}^{(Z)} \\
\partial_t n_\mrm{ion}^{(0)} & = - W^{(1)} n_\mrm{ion}^{(0)}
\label{eq:rate_eq_ion_dens}
\end{split}
\end{equation}
for $Z=1,2,3,\ldots,K$, and the initial neutral density is $n_\mrm{ion}^{(0)}(t=-\infty) = n_\mrm{a}$. The tunnel ionization rate $W^{(Z)}$ in quasi-static approximation creating ions with charge $Z$ is taken from~\cite{Ammosov-1986-Tunnel,PhysRevA.64.013409}. Thus, $W^{(Z)}$ is a function of the modulus of the electric field $\Evec$. The atoms can be at most $K$ times ionized and thus $W^{(K+1)}=0$. The electron density is determined by the ion densities
\begin{equation}
n_\mrm{e} = \sum_{Z} Z n_\mrm{ion}^{(Z)}\,\mbox{.}
\label{eq:el_dens_expl}
\end{equation}
The electron energy density $\mathcal{E}=n_\mrm{e}E_{\mrm{elec}}$ is governed by
\begin{equation}
\partial_t \mathcal{E}=\Jvec\cdot\Evec\,\mbox{.}
\end{equation}
The loss current accounting for ionization losses in the laser field reads
\begin{align}
\Jvec_\mrm{loss} = \frac{\Evec}{|\Evec|^2}\sum_Z I_\mrm{p}^{Z} W^{(Z)} n_\mrm{ion}^{(Z-1)}\,\mbox{,} 
\end{align}
where $I_\mrm{p}^{Z}$ is the ionization potential for creation of an ion with charge $Z$. Even though ionization losses as well as higher order ionization ($Z=2,3,\ldots$) are negligible in the framework of the present study, both are kept for completeness. 

The model is implemented in the code ARCTIC, that solves Eqs.~(\ref{eq:Far_1})-(\ref{eq:Amp_1}) by means of the Yee scheme~\cite{Yee}. We have previously benchmarked the code ARCTIC by the PIC code OCEAN~\cite{PhysRevE.87.043109} accounting for full kinetics of the plasma in \cite{Thiele2017}.

\section{Plasmonic resonances in the plasma slab model}
\label{sub:plasma_slab_mod}
To elaborate the origin of the \change{laser-polarization-dependence} of the THz-emission spectra from elliptically shaped 2C-laser-induced plasmas, we consider a simplified system as sketched in Fig.~5(a) of the main article: A plasma slab of thickness $d$ in $x$ direction with time-invariant electron density $n_0$ and collision frequency $\nu_\mrm{ei}$. Both quantities are translationally invariant in $y$ and $z$. Above and below the slab we assume semi-infinite vacuum with \change{a} constant \change{(relative)} permittivity $\epsilon^\mrm{v}=1$. 
By writing the total electric field as the sum of the laser field $\Evec_\mrm{L}$ and the field due to laser-plasma interaction $\tilde{\Evec}$, Eq.~(2) in the main article reads as
\begin{equation}
\partial_t \Jvec + \nu_\mrm{ei}\Jvec=\frac{q_\mrm{e}^2n_0}{m_\mrm{e}}\tilde{\Evec}+\bm\iota\label{eq:current}
\end{equation}
with the source term
\begin{equation}
\bm\iota = \frac{q_\mrm{e}^2n_\mrm{e}}{m_\mrm{e}}\Evec_\mrm{L}\,\mrm{.}\label{eq:cur_source}
\end{equation}
It is important to note that we assume that the source term $\bm\iota$ depends on the product of the laser electric field and the \change{time-dependent} electron density $n_\mrm{e}(t)$, as it is produced during the laser gas interaction. Only in the description of the response of the plasma slab we make the simplification of a \change{time-invariant} density $n_0$.

Let us now use Eq.~(\ref{eq:current}) and Maxwell's equations to determine the response of the system. In frequency space~(see Sec.~\ref{app:FT} for definition), they can be rewritten for angular frequency $\omega\neq0$ as
\begin{align}
\nabla\times\hat{\tilde{\Evec}}&=\rmi\omega\hat{\tilde{\Bvec}}\label{eq:Faraday_omega}\\
\nabla\times\hat{\tilde{\Bvec}}&=-\rmi\frac{\omega}{c^2}\epsilon\hat{\tilde{\Evec}}+\hat{\Qvec}\label{eq:Ampere_omega}\,\mrm{,}
\end{align}
where for sake of brevity we introduced the source term
\begin{equation}
\hat{\Qvec}=\frac{\mu_0 \hat{\boldsymbol{\iota}}}{-\rmi\omega+\nu_\mrm{ei}}\,\mrm{.}
\end{equation}
The dielectric permittivity $\epsilon$ reads ${\epsilon=\epsilon^\mrm{p}}$ in the plasma slab ($|x|\leq d/2$), and $\epsilon=\epsilon^\mrm{v}=1$ in the vacuum ($|x|> d/2$). The complex dielectric permittivity of the plasma is given by
\begin{equation}
\epsilon^\mrm{p} = 1-\frac{\omega_\mrm{p}^2}{\omega^2+\rmi\omega\nu_\mrm{ei}}\,\mrm{,}
\end{equation}
where the plasma frequency $\omega_\mrm{p}=\sqrt{\frac{n_0q_\mrm{e}^2}{m_\mrm{e}\epsilon_0}}$ involves the time independent density $n_0$ of the slab.
\change{The time-independent collision frequency $\nu_\mrm{ei}$ can be disregarded~($\nu_\mathrm{ei} = 0$) and is kept here just for completeness.}

Same as for $\epsilon$, we consider an excitation that is translational invariant in $y$, that is, $\partial_y\hat{\bm\iota}=0$ and $\partial_y\hat{\Qvec}=0$. Therefore, we can set all the $y$-derivatives to zero and Eqs.~(\ref{eq:Faraday_omega})-(\ref{eq:Ampere_omega}) separate into two sets of equations. The translational invariance of the slab in $z$ allows to write down these two sets of equations in the spatial Fourier domain with respect to $z$~($\partial_z\rightarrow\rmi k_z$) giving
\begin{align}
\partial_x \breve{\tilde{E}}_y&=\rmi\omega\breve{\tilde{B}}_z\label{eq:TE_1}\\
\mathrm{(S)}\qquad\qquad\qquad-\rmi k_z\breve{\tilde{E}}_y&=\rmi\omega\breve{\tilde{B}}_x\label{eq:TE_2}\\
\rmi k_z\breve{\tilde{B}}_x-\partial_x\breve{\tilde{B}}_z&=-\rmi\frac{\omega}{c^2}\epsilon\breve{\tilde{E}}_y+\breve{Q}_y\label{eq:TE_3}
\end{align}
and
\begin{align} 
\rmi k_z\breve{\tilde{E}}_x-\partial_x\breve{\tilde{E}}_z&=\rmi\omega\breve{\tilde{B}}_y\label{eq:TM_1}\\
\mathrm{(P)}\qquad\qquad\qquad-\rmi k_z \breve{\tilde{B}}_y&=-\rmi\frac{\omega}{c^2}\epsilon\breve{\tilde{E}}_x+\breve{Q}_x\label{eq:TM_2}\\
\partial_x\breve{\tilde{B}}_y&=-\rmi\frac{\omega}{c^2}\epsilon\breve{\tilde{E}}_z+\breve{Q}_z \mrm{,}\label{eq:TM_3}
\end{align}
where ``$\,\,\breve{}\,\,$'' indicates the $(\omega,k_z)$-domain according to the definition in Sec.~\ref{app:FT}. The \first set of equations (\ref{eq:TE_1})-(\ref{eq:TE_3}) \blue{is associated with the so-called S polarization. In the waveguide context, it is also often termed the transverse electric (TE) mode}, because the only electric field component $E_y$ is polarized in the transverse translational invariant direction. Here, the only fields different from zero are $(\breve{\tilde{B}}_x,\breve{\tilde{E}}_y,\breve{\tilde{B}}_z)$. The \second set of equations (\ref{eq:TM_1})-(\ref{eq:TM_3}) \blue{is associated with the so-called P polarization} and describes the evolution of $(\breve{\tilde{E}}_x,\breve{\tilde{B}}_y,\breve{\tilde{E}}_z)$. \blue{In the waveguide context it is also often termed the transverse magnetic (TM) mode.}

In the following, we will consider two different configurations: 
\begin{itemize}
	\item[(i)] \blue{S polarization} with transverse excitation in $y$~($\breve{\iota}_y\neq0\neq \breve{Q}_y$ and $\breve{Q}_x=0= \breve{Q}_z$)
	\item[(ii)] \blue{P polarization} with transverse excitation in $x$~($\breve{\iota}_x\neq0\neq \breve{Q}_x$ and $\breve{Q}_z=0= \breve{Q}_y$)
\end{itemize}
Note that (i) corresponds to the THz generation by $y$-polarized \change{"elliptical beams"} and (ii) by $x$-polarized \change{"elliptical beams"} as investigated in the main article. %In the following, the Maxwell's equations Eqs.~(\ref{eq:TE_1})-(\ref{eq:TM_3}) for the plasma slab are solved for these two cases.

Before computing the response of the plasma slab to an excitation by $\bm{\iota}$, it is interesting to investigate the resonances of the homogeneous system~($\bm{\iota}=0$).
To this end, it is convenient to solve the reflection transmission problem and to consider the reflection coefficient. Singularities of the reflection coefficient describe resonant modes, i.e., modes which provide non-zero fields in the slab without an incoming wave and without a source term. 	When then exciting the system by the source $\bm{\iota}$, these modes are most likely excited and thus of great interest.
%To imagine another situation of a similar kind, think of a harmonic oscillator which provides homogeneous solutions oscillating with the resonance frequency. Those are nothing but resonant modes. When considering the inhomogeneous case, the harmonic oscillator responds strongly when exciting the system close to the resonance frequency because of these resonant modes.}

\subsection{Reflection coefficient of a slab}

\begin{figure}[b]
	\centering
	\includegraphics[width=0.8\columnwidth]{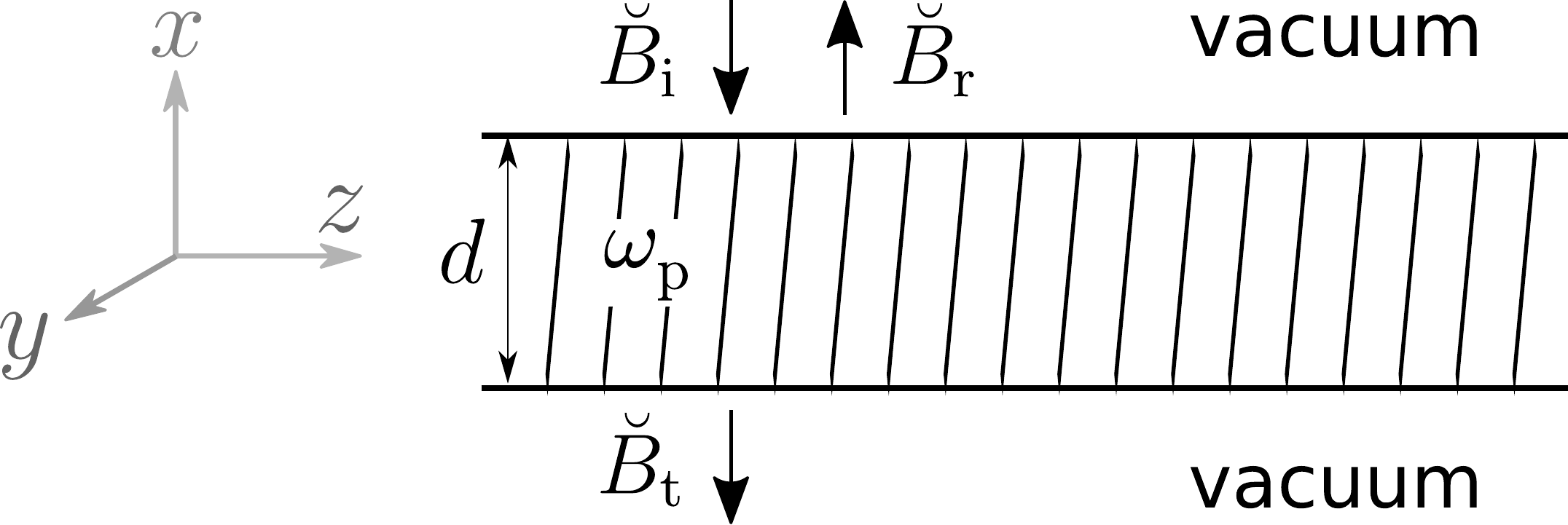}
	\caption{Illustration of the plasma slab. \change{For the magnetic field components, see text.}}
	\label{fig:slab}
\end{figure}

In the following, we consider the problem \change{of computing} the reflection coefficient for a plane wave arriving from one of the vacuum half-spaces and interacting with a plasma slab~(see Fig.~\ref{fig:slab}). \change{To this end}, we will operate in the $(\omega,k_z)$-space. We will \change{detail} the \blue{P polarization} case, because it is of most interest. The \blue{S polarization} case follows analogously with a small modification which will \change{be} highlighted in the coming derivation.

We denote by $\breve{B}_\mrm{i}$ the amplitude of the magnetic field of the incident \blue{P polarized} wave and by $\breve{B}_\mrm{r}$ the amplitude of the reflected wave. Without loss of generality, the incident wave arrives from the positive half-space. Then, the magnetic field in the vacuum for $x\geq d/2$ writes
\begin{equation}
\change{\breve{B}^{\rm v}_y(x)=\breve{B}_\mrm{i}\exp\left[-\Lambda^\mrm{v}\left(x-\frac{d}{2}\right)\right]+\breve{B}_\mrm{r}\exp\left[\Lambda^\mrm{v}\left(x-\frac{d}{2}\right)\right]}\,\mrm{,}
\end{equation}
\change{where $\Lambda^\mrm{v}=\sqrt{k_z^2 - \omega^2/c^2}$}. Denoting its $x$-derivative by $\breve{B}_y^\prime=\partial_x\breve{B}_y$, this implies the relation
\begin{equation}
\change{\left.\left(
	\begin{matrix}
	\breve{B}^{\rm v}_y \\ \breve{B}_y^{\rm v\prime}
	\end{matrix}
	\right)\right|_{x=\frac{d}{2}}}
=
\underbrace{
	\left(
	\begin{matrix}
	1 & 1 \\ -\Lambda^\mrm{v} & \Lambda^\mrm{v}
	\end{matrix}
	\right)	
}_{=:\mathbf{K}}
\left(
\begin{matrix}
\breve{B}_\mrm{i} \\ \breve{B}_\mrm{r}
\end{matrix}
\right)
\,\mbox{,}
%\label{eq:matrix_R}
\end{equation}
where we introduced the transformation matrix $\mathbf{K}$.

For \blue{P polarization}, only the transverse field $\breve{B}_y$ is continuous at an interface while for the derivative we can only use that $\epsilon^{-1}\partial_x\breve{B}_y$ is continuous. The situation is easier for \blue{S polarization} where both $\breve{E}_y$ and $\breve{E}_y^\prime$ are continuous at an interface. Thus, to handle the \blue{P polarization} and later on the \blue{S polarization} case in the same manner, we introduce $\alpha_\mrm{P}=1/\epsilon^\mrm{p}$ and $\alpha_\mrm{S}=1$. Then, we relate the fields at the vacuum plasma interface at $x=d/2$ by
\begin{equation}
\left.\left(
\begin{matrix}
\breve{B}^{\rm p}_y \\ \breve{B}_y^{\rm p\prime}
\end{matrix}
\right)\right|_{x=\frac{d}{2}}
=
\underbrace{
	\left(
	\begin{matrix}
	1 & 0 \\ 0 & \alpha_\mrm{P}^{-1}
	\end{matrix}
	\right)	
}_{=:\mathbf{T}}
\left.\left(
\begin{matrix}
\breve{B}^{\rm v}_y \\ \breve{B}_y^{\rm v \prime}
\end{matrix}
\right)\right|_{x=\frac{d}{2}}
\,\mbox{,}
\label{eq:matrix_T}
\end{equation}
where we introduced the interface transition matrix $\mathbf{T}$. 

The magnetic field in the plasma~($|x|<d/2$) can be decomposed \change{into a forward $\breve{B}_\mrm{+}$ and a backward $\breve{B}_\mrm{-}$ running wave as}
\begin{equation}
\breve{B}^{\rm p}_y(x)=\breve{B}_\mrm{+}\exp\left(\Lambda^\mrm{p}x\right)+\breve{B}_\mrm{-}\exp\left(-\Lambda^\mrm{p}x\right)\,\mrm{,}
\end{equation}
\change{where $\Lambda^\mrm{p}=\sqrt{k_z^2 - \epsilon^\mrm{p}\omega^2/c^2}$}. Using this equation, a short computation relates fields at the two interfaces inside the plasma
\begin{equation}
\left.\left(
\begin{matrix}
\breve{B}^{\rm p}_y \\ \breve{B}_y^{\rm p \prime}
\end{matrix}
\right)\right|_{x=-\frac{d}{2}}
=\mathbf{M}
\left.\left(
\begin{matrix}
\breve{B}^{\rm p}_y \\ \breve{B}_y^{\rm p \prime}
\end{matrix}
\right)\right|_{x=\frac{d}{2}}
\,\mbox{.}
%\label{eq:matrix_R}
\end{equation}
with
\begin{equation}
\mathbf{M}=
\left(
\begin{matrix}
\cosh\left(\Lambda^\mrm{p}d\right) & \sinh\left(\Lambda^\mrm{p}d\right)/\Lambda^\mrm{p} \\ \Lambda^\mrm{p}\sinh\left(\Lambda^\mrm{p}d\right) & \cosh\left(\Lambda^\mrm{p}d\right)
\end{matrix}
\right)\,\mrm{.}
\end{equation}
In complete analogy to Eq.~(\ref{eq:matrix_T}), we obtain the fields in vaccum at the lower interface at $x=-d/2$ by
\begin{equation}
\left.\left(
\begin{matrix}
\breve{B}^{\rm v}_y \\ \breve{B}_y^{\rm v \prime}
\end{matrix}
\right)\right|_{x=-\frac{d}{2}}
=
\mathbf{T}^{-1}
\left.\left(
\begin{matrix}
\breve{B}^{\rm p}_y \\ \breve{B}_y^{\rm p \prime}
\end{matrix}
\right)\right|_{x=-\frac{d}{2}}
\,\mbox{.}
%\label{eq:matrix_R}
\end{equation}
We denote by $\breve{B}_\mrm{t}$ the amplitude of the transmitted wave. Because there is no incoming wave from the negative half-space, the field in the vacuum for $x\leq d/2$ reads
\begin{equation}
\breve{B}^{\rm v}_y(x)=\breve{B}_\mrm{t}\exp\left[-\Lambda^\mrm{v}\left(x-\frac{d}{2}\right)\right]\,\mrm{.}
\end{equation}
In summary, we therefore obtain
\begin{align}
\breve{B}_\mrm{t}
\left(
\begin{matrix}
1 \\ -\Lambda^\mrm{v}
\end{matrix}
\right)
=
\left.\left(
\begin{matrix}
\breve{B}^{\rm v}_y \\ \breve{B}_y^{\rm v \prime}
\end{matrix}
\right)\right|_{x=-\frac{d}{2}}
= 
\mathbf{T}^{-1}\mathbf{M}\mathbf{T}\mathbf{K}
\left(
\begin{matrix}
\breve{B}_\mrm{i} \\ \breve{B}_\mrm{r}
\end{matrix}
\right)\mrm{.}	
\end{align}
After performing all the matrix multiplications, we end up two equations relating the field amplitudes $\breve{B}_\mrm{i}$, $\breve{B}_\mrm{r}$, and $\breve{B}_\mrm{t}$. We can thus eliminate the transmitted field amplitude $\breve{B}_\mrm{t}$ and get the complex valued reflection coefficient
\begin{equation}
r_{\rm P}:=\frac{\breve{B}_\mrm{r}}{\breve{B}_\mrm{i}} = \frac{{\Lambda^{\rm v}}^2-{\Lambda^{\rm p}}^2\alpha_{\rm P}^2}
{{\Lambda^{\rm v}}^2+{\Lambda^{\rm p}}^2\alpha_{\rm P}^2
	+\frac{2{\Lambda^{\rm v}} {\Lambda^{\rm p}}\alpha_{\rm P}}{\tanh\!\left({\Lambda^{\rm p}}d\right)}}\mrm{.}
\end{equation}
This derivation can be straightforwardly repeated for the \blue{S polarization} case, where we govern $\breve{E}_y$ instead of $\breve{B}_y$ and just have to replace $\alpha_{\rm P}$ by $\alpha_{\rm S}$. Then, we obtain the intensity \change{reflectivity} $\rho_{\rm S/P}:= |r_{\rm S/P}|^2$ as
\begin{equation}
\rho_{\rm S/P}= \left|\frac{{\Lambda^{\rm v}}^2-{\Lambda^{\rm p}}^2\alpha_{\rm S/P}^2}
{{\Lambda^{\rm v}}^2+{\Lambda^{\rm p}}^2\alpha_{\rm S/P}^2
	+\frac{2{\Lambda^{\rm v}} {\Lambda^{\rm p}}\alpha_{\rm S/P}}{\tanh\!\left({\Lambda^{\rm p}}d\right)}}\right|^2. \label{eq:rho_slab} 
\end{equation}
\change{This is exactly Eq. (2) of the main article.}

\subsection{S polarization with transverse excitation in $y$}
\label{par:TE}
In the following, we consider the full inhomogeneous set of Eqs.~(\ref{eq:TE_1})-(\ref{eq:TM_3}) including the excitation source term $\bm{\iota}$. While in the general case these equations have to be solved numerically, for the plasma slab they can be solved even analytically as \change{it} will be shown in the following.

Let us start with case (i), that is, $\breve{\bm\iota}=\breve{\iota}_y\evec_y$. Firstly, the general solutions inside the plasma slab and the neutral gas are computed, and secondly, the continuity of the transverse fields is used to determine the entire solution.

Equations~(\ref{eq:TE_1})-(\ref{eq:TE_3}) give the evolution equation for the transverse field $\breve{\tilde{E}}_y$ inside the plasma and the neutral gas
\begin{equation}
\partial_x^2 \breve{\tilde{E}}_y - \left(k_z^2 - \frac{\omega^2}{c^2}\epsilon\right) \breve{\tilde{E}}_y = -\rmi \omega \breve{Q}_y\,\mrm{.}\label{eq:wave_TE}
\end{equation}
For sake of simplicity, we consider that $\breve{Q}_y$ is constant inside the plasma slab and zero outside of it. In particular, this choice makes $\breve{Q}_y$ symmetric with respect to $x$, and $\breve{\tilde{E}}_y$ has to obey the same symmetry. Then, the general symmetric solution in the plasma ($|x|\le d/2$) reads
\begin{equation}
\breve{\tilde{E}}_y^\mrm{p} = A^\mrm{p} \cosh\left(\Lambda^\mrm{p}x\right)-\frac{\rmi \omega \breve{Q}_y}{(\Lambda^\mrm{p})^2}\left[\cosh\left(\Lambda^\mrm{p}x\right)-1\right]\,\mrm{.}
\end{equation}
In the vacuum at $x>d/2$ the general solution reads
\begin{equation}
\breve{\tilde{E}}_y^\mrm{v} = A^\mrm{v}\exp\left[\mp\Lambda^\mrm{v}\left(x-\frac{d}{2}\right)\right]\,\mrm{.}\label{eq:Eyv}
\end{equation}
For $k_z^2\geq\omega^2/c^2$, we use the ``$-$'' sign since for the ``$+$'' sign the field would grow exponentially when $x\rightarrow\infty$. For $k_z^2<\omega^2/c^2$, we use the ``$+$'' sign to obtain only outgoing propagating waves (along $x$). Note that solution in the vaccum at $x<d/2$ follows from symmetry considerations. 

According to Maxwell's interface conditions $\breve{\tilde{E}}_y$, $\breve{\tilde{B}}_z$ and thus $\partial_x \breve{\tilde{E}}_y$ have to be continuous at the plasma-vacuum interface. These two conditions determine the yet unknown amplitudes $A^\mrm{p}$ and $A^\mrm{v}$ to
\begin{align}
A^\mrm{p} &= \frac{\rmi\omega\breve{Q}_y}{(\Lambda^\mrm{p})^2}\left(\frac{\Lambda^\mrm{v}}{D^\mrm{p}}+1\right) \\
A^\mrm{v} &= \frac{\rmi\omega\breve{Q}_y}{(\Lambda^\mrm{p})^2}\left[\frac{\Lambda^\mrm{v}}{D^\mrm{p}}\cosh\left(\frac{\Lambda^\mrm{p}d}{2}\right)+1\right]\,\mrm{,} 
\end{align} 
with common denominator
\begin{equation}
D^\mrm{p} = \mp\Lambda^\mrm{p}\sinh\left(\frac{\Lambda^\mrm{p}d}{2}\right)-\Lambda^\mrm{v}\cosh\left(\frac{\Lambda^\mrm{p}d}{2}\right)\,\mrm{.}
\end{equation}
Finally, Eqs.~(\ref{eq:TE_1})-(\ref{eq:TE_2}) determine the magnetic field components as
\begin{align}
\breve{\tilde{B}}_x&=-\frac{k_z}{\omega}\breve{\tilde{E}}_y\\
\breve{\tilde{B}}^\mrm{p}_z&= \frac{ \breve{Q}_y}{\Lambda^\mrm{p}}\frac{\Lambda^\mrm{v}}{D^\mrm{p}}\sinh\left(\Lambda^\mrm{p}x\right) \\
\breve{\tilde{B}}^\mrm{v}_z&= \mp\frac{\Lambda^\mrm{v}}{\rmi\omega}A^\mrm{v}\exp\left[\mp\Lambda^\mrm{v}\left(x-\frac{d}{2}\right)\right] \,\mrm{.}\label{eq:Bzv}
\end{align}
\change{Equations (\ref{eq:Eyv}) and (\ref{eq:Bzv}) were used in Sec. 3 of the main article to compute the transverse Poynting fluxes via Eq.~(\ref{eq:S}).}

\subsection{P polarization case with transverse excitation in $x$}
\label{par:trans_TM}
Next, case (ii) with $\breve{\bm\iota}=\breve{\iota}_x\evec_x$ is considered. Equations.~(\ref{eq:TM_1})-(\ref{eq:TM_3}) give the evolution equation for the transverse field $\breve{\tilde{B}}_y$ inside the plasma and neutral gas respectively,
\begin{equation}
\partial_x^2 \breve{\tilde{B}}_y - \Lambda^2 \breve{\tilde{B}}_y = -\rmi k_z \breve{Q}_x\,\mrm{.}\label{eq:wave_TM_1}
\end{equation}
In analogy to the \blue{S polarization} case in the previous section, we obtain in the plasma
\begin{equation}
\breve{\tilde{B}}_y^\mrm{p} = A^\mrm{p} \cosh\left(\Lambda^\mrm{p}x\right)-\frac{\rmi k_z \breve{Q}_x}{(\Lambda^\mrm{p})^2}\left[\cosh\left(\Lambda^\mrm{p}x\right)-1\right]
\end{equation}
and in upper vacuum
\begin{equation}
\breve{\tilde{B}}_y^\mrm{v} = A^\mrm{v}\exp\left[\mp\Lambda^\mrm{v}\left(x-\frac{d}{2}\right)\right]\,\mrm{.}\label{eq:Byv}
\end{equation}

The difference to the \blue{S polarization} case appears when applying the interface conditions at the plasma-air interface: $\breve{\tilde{B}}_y$ and $\breve{\tilde{E}}_z$ are continuous but according to Eq.~(\ref{eq:TM_3}) $\partial_x\breve{\tilde{B}}_y$ is not, because $\epsilon$ changes at the interface. Applying these \blue{P polarization} interface conditions determines $A^\mrm{p}$ and $A^\mrm{v}$ to
\begin{align}
A^\mrm{p} &= \frac{\rmi k_z \breve{Q}_x}{(\Lambda^\mrm{p})^2}\left(\frac{\Lambda^\mrm{v}}{D^\mrm{p}}+1\right) \\
A^\mrm{v} &= \frac{\rmi k_z \breve{Q}_x}{(\Lambda^\mrm{p})^2}\left[\frac{\Lambda^\mrm{v}}{D^\mrm{p}}\cosh\left(\frac{\Lambda^\mrm{p}d}{2}\right)+1\right]\,\mrm{,} 
\end{align} 
with common denominator
\begin{equation}
D^\mrm{p} = \mp\frac{1}{\epsilon^\mrm{p}}\Lambda^\mrm{p}\sinh\left(\frac{\Lambda^\mrm{p}d}{2}\right)-\Lambda^\mrm{v}\cosh\left(\frac{\Lambda^\mrm{p}d}{2}\right)\,\mrm{.}
\end{equation}
Finally, Eqs.~(\ref{eq:TM_2})-(\ref{eq:TM_3}) determine the electric field components as
\begin{align}
\breve{\tilde{E}}_x&=\frac{k_z c^2}{\omega\epsilon}\breve{\tilde{B}}_y-\frac{\rmi c^2}{\omega\epsilon}\breve{Q}_x\\
\breve{\tilde{E}}^\mrm{p}_z&=-\frac{k_z \breve{Q}_xc^2}{\omega\epsilon^\mrm{p}\Lambda^\mrm{p}}\frac{\Lambda^\mrm{v}}{D^\mrm{p}}\sinh\left(\Lambda^\mrm{p}x\right)\\
\breve{\tilde{E}}^\mrm{v}_z&=\mp\frac{\rmi c^2\Lambda^\mrm{v}}{\omega}A^\mrm{v}\exp\left[\mp\Lambda^\mrm{v}\left(x-\frac{d}{2}\right)\right]\,\mrm{.}\label{eq:Ezv}
\end{align}
\change{Equations (\ref{eq:Byv}) and (\ref{eq:Ezv}) were used in Sec. 3 of the main article to compute the transverse Poynting fluxes via Eq.~(\ref{eq:S}).}

\section{THz emission from elliptically shaped \\ microplasmas}
\label{app:ne}

In the main article we investigate terahertz~(THz) emission from elliptically shaped two-color~(2C)-laser-induced gas-plasmas. 
In such a configuration, the free electron density profile with a small transverse size along $x$ and a large transverse size along $y$ is created. 
%As can be seen from Fig.~\ref{fig:Ey_n0_qTE}, 
For the considered microplasmas, the transverse plasma profile is strongly elliptical, that is, along $x$ direction the plasma size is less than 1~$\mu$m, whereas along $y$ direction the plasma is approximately 10~$\mu$m wide. 
%As can be seen from the transverse electron density profiles~(bright white lines), the electron density gradients are large along $x$ and significantly smaller along $y$. 
Thus, by rotating the linear laser electric field polarization with respect to the transverse plasma profile, we can select the strength of the electron density gradients along the excitation direction. %It is clear that the excitation and thus the displacement of the electrons along $x$ will lead to a stronger charge separation between the electrons and ions than the excitation along $y$.

\begin{figure}[t]
	\centering
	\includegraphics[width=1.0\columnwidth]{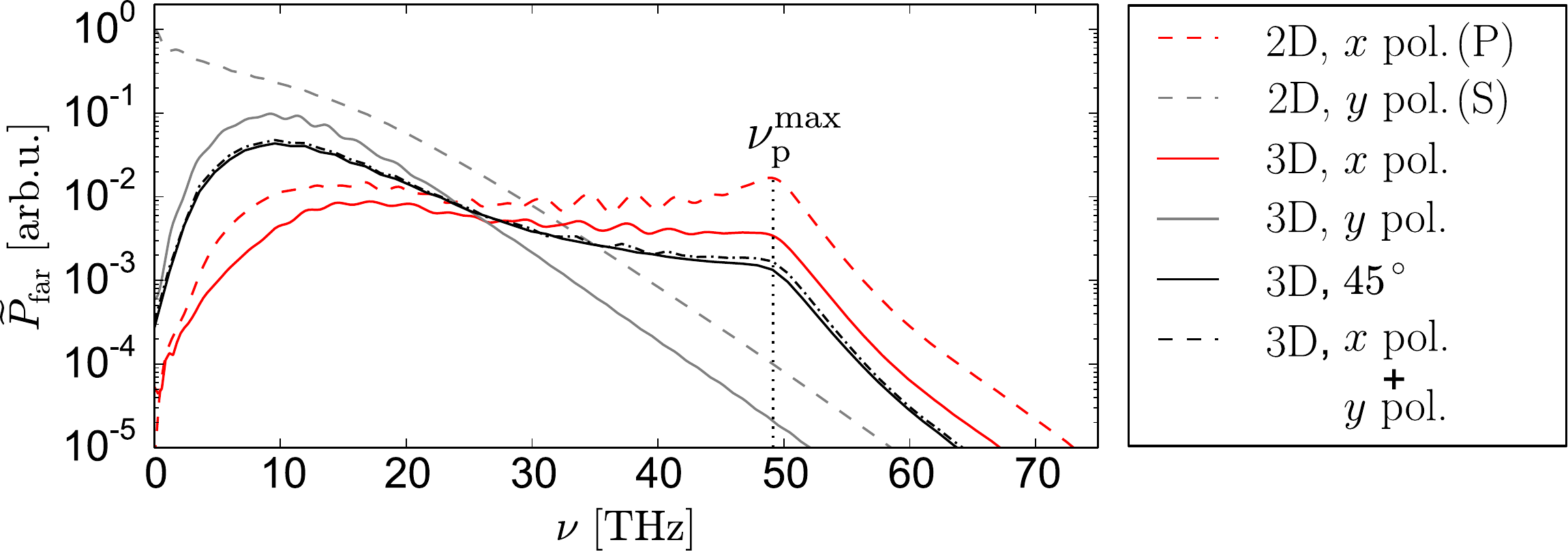}
	\caption{Angularly integrated far-field spectra for the elliptical beams from Sec.~4 of the main article and corresponding results from 2D simulations assuming translational invariance in $y$ (dashed lines). The solid black line specifies the emission spectrum from a 3D simulation with laser polarization at $45^\circ$ in the $xy$ plane. The \change{black dash-dotted} line is computed from the superimposed fields \blue{obtained with an $x$- and a $y$-polarized laser} in 3D.}
	\label{fig:spectra_elipt}
\end{figure}

As \change{it} can be seen in the 3D angularly integrated far-field spectra presented in Fig.~\ref{fig:spectra_elipt}, when exciting the plasma along strong electron density gradients~\blue{($x$-polarized laser associated with P polarization)}, the THz spectrum is broadened up to about 50~THz~\blue{(dark red solid line)}, which corresponds to the maximum plasma frequency $\nu_\mrm{p}^\mrm{max}$. In contrast, when exciting the plasma along the weak electron density gradients~\blue{($y$-polarized laser associated with S polarization)} no such broadening is found~\blue{(light gray solid line)}. 

%Let us give a simple explanation for the difference in the THz spectra for qTE and qTM polarization: When the plasma current is excited along the direction of weak plasma gradients, the charge separation created due to the displacement of the electrons with respect to the ions is relatively small. Thus, the restoring force is small and the electron current can persist for a long time which leads to a quasi-DC current resulting in the emission of rather low frequencies. In contrast, when the plasma current is excited along the direction of strong plasma gradients, a significant charge separation force pulls the electrons back to the ions resulting in higher frequency components in the current and the emitted radiation. 

Results of corresponding 2D simulations~($\partial_y=0$) are shown as dashed lines in Fig.~\ref{fig:spectra_elipt}. Here, we find a similar behavior as in 3D: no broadening if the laser electric field is oriented in the now translationally invariant $y$ direction (\blue{S}), and broadening up to $\nu_\mrm{p}^\mrm{max}$ if the laser electric field points in the direction of the strong electron density gradient, that is, along the $x$ direction (\blue{P}). 
Treating the problem in 2D geometry, i.e., assuming translational invariance in one transverse direction (here $y$), the electromagnetic fields separate into two cases:
%The advantage of treating the problem in 2D geometry is that by assuming translational invariance in one transverse direction (here $y$) the electromagnetic fields separate into two cases, which simplifies any analytical approach drastically: 
\blue{S polarization case} that governs the fields $B_{x}$, $E_{y}$, $B_{z}$ for a $y$-polarized driving laser pulse and the \blue{P polarization case} that governs the fields $E_{x}$, $B_{y}$, $E_{z}$ for an $x$-polarized driving laser pulse. Any other polarization state in 2D can be written as the superposition of these two cases\change{, if coupling through the laser generated plasma is ignored.} For example, an incoming laser pulse that is linearly polarized under $45^\circ$ in the $xy$ plane will give an electric field solution that can be written as \blue{$\Evec=\Evec^\mrm{S}/\sqrt{2}+\Evec^\mrm{P}/\sqrt{2}$, where $\Evec^\mrm{S}$ and $\Evec^\mrm{P}$ are the solutions for an S and a P polarized driving laser pulse, respectively.} Therefore, if the 3D elliptical beam can be indeed approximated by the idealized 2D case, and no detrimental nonlinear coupling occurs, this property should hold. We checked this by comparing the angularly integrated THz-far-field power spectrum for a simulation with polarization at $45^\circ$~(black solid line) and the result for the superposed fields~(black dashed line) in Fig.~\ref{fig:spectra_elipt}. Both overlap almost perfectly, \change{which further justifies our} analysis of the 2D configuration. Moreover, the possibility of superposing \blue{the THz fields which are produced by an $x$- and $y$-polarized laser pulse} could be important for applications, because it implies that the THz emission spectrum can be tuned by rotating the linear polarization of the incoming \change{elliptical} laser pulse.

	\end{appendix}

	\bibliography{mybibfile}
	
\end{document}